\documentstyle[10pt,epsf,epsfig,dp_delphititle,float,hangcaption,xspace,
amssymb,amsfonts,amsmath,amsthm,cite,graphicx]{dp_delphi}
\makeindex
\def\DpPaperGroup{PH-EP}
\def\DpPaperRef{2006-037}
\def\DpDate{22 November 2006}
\def\DpAuthors{DELPHI Collaboration}
\def\DpSubmit{(Accepted by Eur. Phys. J. C)}
\def\DpTitle{{Investigation of \\
 Colour Reconnection in WW Events with
 the DELPHI detector at LEP-2 }}
\def\DpComment{}
\def\DpEMail{}

\newcommand{\PLB}[3]{{\em Phys. Lett.} {\bf B}{\bf{#1} }{(#2) }{#3}}
\newcommand{\PRL}[3]{{\em Phys. Rev. Lett.} {\bf{#1} }{(#2) }{#3}}
\newcommand{\PRD}[3]{{\em Phys. Rev.} {\bf D}{\bf{#1} }{(#2) }{#3}}
\newcommand{\ZPC}[3]{{\em Z. Phys.} {\bf C}{\bf{#1} }{(#2) }{#3}}

\newcommand{\CPC}[3]{{\em Comp. Phys. Comm.} {\bf{#1} }{(#2) }{#3}}

\begin{document}
\makeatletter
\makeatother

\begin{titlepage}
\pagenumbering{roman}

\CERNpreprint{\DpPaperGroup}{\DpPaperRef}   
\date{{\small\DpDate}}                      
\title{\DpTitle}                            
\address{\DpAuthors}                        

\begin{shortabs}                            
\noindent
%
\noindent

In the reaction
$\mathrm{e^+e^- \rightarrow WW \rightarrow (q_1 \bar q_2)
(q_3 \bar q_4)}$ the usual hadronization models treat the
colour singlets $\mathrm{q_1 \bar q_2}$ and $\mathrm{q_3 \bar q_4}$ coming from two
W bosons independently.
However, since the final state partons
 may coexist in space and time, cross-talk between the two evolving
 hadronic systems may be possible during fragmentation through soft
 gluon exchange. This effect is known as Colour Reconnection. In this
 article the results of the investigation of Colour Reconnection effects
 in fully hadronic decays of W pairs in DELPHI at LEP are presented.
 Two complementary analyses were performed, studying the particle flow
 between  jets and 
 W mass estimators,
 with negligible correlation between them, and the results were combined
 and compared to models. In the framework of the SK-I model, the
 value for its $\kappa$ parameter most compatible with the data was
 found to be:
\begin{equation*}
\kappa_{\mathrm{SK-I}}= 2.2^{+2.5}_{-1.3}
\end{equation*}
\noindent corresponding to the probability of
reconnection ${\cal P}_{\mathrm{reco}}$ to be in the range
$\mathrm{0.31 < {\cal P}_{reco} < 0.68}$  at 68\% confidence level
with its best value at 0.52.

\end{shortabs}

\vfill

\begin{center}
\DpSubmit \ \\          
\DpComment \ \\
\DpEMail \ \\
\end{center}

\vfill
\clearpage

\headsep 10.0pt

\addtolength{\textheight}{10mm}
\addtolength{\footskip}{-5mm}
\begingroup
%
\newcommand{\DpName}[2]{\hbox{#1$^{\ref{#2}}$},\hfill}
\newcommand{\DpNameTwo}[3]{\hbox{#1$^{\ref{#2},\ref{#3}}$},\hfill}
\newcommand{\DpNameThree}[4]{\hbox{#1$^{\ref{#2},\ref{#3},\ref{#4}}$},\hfill}
\newskip\Bigfill \Bigfill = 0pt plus 1000fill
\newcommand{\DpNameLast}[2]{\hbox{#1$^{\ref{#2}}$}\hspace{\Bigfill}}

%
\footnotesize
\noindent
\DpName{J.Abdallah}{LPNHE}
\DpName{P.Abreu}{LIP}
\DpName{W.Adam}{VIENNA}
\DpName{P.Adzic}{DEMOKRITOS}
\DpName{T.Albrecht}{KARLSRUHE}
\DpName{R.Alemany-Fernandez}{CERN}
\DpName{T.Allmendinger}{KARLSRUHE}
\DpName{P.P.Allport}{LIVERPOOL}
\DpName{U.Amaldi}{MILANO2}
\DpName{N.Amapane}{TORINO}
% ...
\normalsize
\endgroup
\newpage

\titlefoot{Department of Physics and Astronomy, Iowa State
     University, Ames IA 50011-3160, USA
    \label{AMES}}
\titlefoot{IIHE, ULB-VUB,
     Pleinlaan 2, B-1050 Brussels, Belgium
    \label{BRUSSELS}}
\titlefoot{Physics Laboratory, University of Athens, Solonos Str.
     104, GR-10680 Athens, Greece
    \label{ATHENS}}
\titlefoot{Department of Physics, University of Bergen,
     All\'egaten 55, NO-5007 Bergen, Norway
    \label{BERGEN}}
\titlefoot{Dipartimento di Fisica, Universit\`a di Bologna and INFN,
     Via Irnerio 46, IT-40126 Bologna, Italy
    \label{BOLOGNA}}
\titlefoot{Centro Brasileiro de Pesquisas F\'{\i}sicas, rua Xavier Sigaud 150,
     BR-22290 Rio de Janeiro, Brazil
    \label{BRASIL-CBPF}}
\titlefoot{Inst. de F\'{\i}sica, Univ. Estadual do Rio de Janeiro,
     rua S\~{a}o Francisco Xavier 524, Rio de Janeiro, Brazil
    \label{BRASIL-IFUERJ}}
\titlefoot{Coll\`ege de France, Lab. de Physique Corpusculaire, IN2P3-CNRS,
     FR-75231 Paris Cedex 05, France
    \label{CDF}}
\titlefoot{CERN, CH-1211 Geneva 23, Switzerland
    \label{CERN}}
\titlefoot{Institut de Recherches Subatomiques, IN2P3 - CNRS/ULP - BP20,
     FR-67037 Strasbourg Cedex, France
    \label{CRN}}
\titlefoot{Now at DESY-Zeuthen, Platanenallee 6, D-15735 Zeuthen, Germany
    \label{DESY}}
\titlefoot{Institute of Nuclear Physics, N.C.S.R. Demokritos,
     P.O. Box 60228, GR-15310 Athens, Greece
    \label{DEMOKRITOS}}
\titlefoot{FZU, Inst. of Phys. of the C.A.S. High Energy Physics Division,
     Na Slovance 2, CZ-180 40, Praha 8, Czech Republic
    \label{FZU}}
\titlefoot{Dipartimento di Fisica, Universit\`a di Genova and INFN,
     Via Dodecaneso 33, IT-16146 Genova, Italy
    \label{GENOVA}}
\titlefoot{Institut des Sciences Nucl\'eaires, IN2P3-CNRS, Universit\'e
     de Grenoble 1, FR-38026 Grenoble Cedex, France
    \label{GRENOBLE}}
\titlefoot{Helsinki Institute of Physics and Department of Physical Sciences,
     P.O. Box 64, FIN-00014 University of Helsinki, 
     \indent~~Finland
    \label{HELSINKI}}
\titlefoot{Joint Institute for Nuclear Research, Dubna, Head Post
     Office, P.O. Box 79, RU-101 000 Moscow, Russian Federation
    \label{JINR}}
\titlefoot{Institut f\"ur Experimentelle Kernphysik,
     Universit\"at Karlsruhe, Postfach 6980, DE-76128 Karlsruhe,
     Germany
    \label{KARLSRUHE}}
\titlefoot{Institute of Nuclear Physics PAN,Ul. Radzikowskiego 152,
     PL-31142 Krakow, Poland
    \label{KRAKOW1}}
\titlefoot{Faculty of Physics and Nuclear Techniques, University of Mining
     and Metallurgy, PL-30055 Krakow, Poland
    \label{KRAKOW2}}
\titlefoot{Universit\'e de Paris-Sud, Lab. de l'Acc\'el\'erateur
     Lin\'eaire, IN2P3-CNRS, B\^{a}t. 200, FR-91405 Orsay Cedex, France
    \label{LAL}}
\titlefoot{School of Physics and Chemistry, University of Lancaster,
     Lancaster LA1 4YB, UK
    \label{LANCASTER}}
\titlefoot{LIP, IST, FCUL - Av. Elias Garcia, 14-$1^{o}$,
     PT-1000 Lisboa Codex, Portugal
    \label{LIP}}
\titlefoot{Department of Physics, University of Liverpool, P.O.
     Box 147, Liverpool L69 3BX, UK
    \label{LIVERPOOL}}
\titlefoot{Dept. of Physics and Astronomy, Kelvin Building,
     University of Glasgow, Glasgow G12 8QQ
    \label{GLASGOW}}
\titlefoot{LPNHE, IN2P3-CNRS, Univ.~Paris VI et VII, Tour 33 (RdC),
     4 place Jussieu, FR-75252 Paris Cedex 05, France
    \label{LPNHE}}
\titlefoot{Department of Physics, University of Lund,
     S\"olvegatan 14, SE-223 63 Lund, Sweden
    \label{LUND}}
\titlefoot{Universit\'e Claude Bernard de Lyon, IPNL, IN2P3-CNRS,
     FR-69622 Villeurbanne Cedex, France
    \label{LYON}}
\titlefoot{Dipartimento di Fisica, Universit\`a di Milano and INFN-MILANO,
     Via Celoria 16, IT-20133 Milan, Italy
    \label{MILANO}}
\titlefoot{Dipartimento di Fisica, Univ. di Milano-Bicocca and
     INFN-MILANO, Piazza della Scienza 3, IT-20126 Milan, Italy
    \label{MILANO2}}
\titlefoot{IPNP of MFF, Charles Univ., Areal MFF,
     V Holesovickach 2, CZ-180 00, Praha 8, Czech Republic
    \label{NC}}
\titlefoot{NIKHEF, Postbus 41882, NL-1009 DB
     Amsterdam, The Netherlands
    \label{NIKHEF}}
\titlefoot{National Technical University, Physics Department,
     Zografou Campus, GR-15773 Athens, Greece
    \label{NTU-ATHENS}}
\titlefoot{Physics Department, University of Oslo, Blindern,
     NO-0316 Oslo, Norway
    \label{OSLO}}
\titlefoot{Dpto. Fisica, Univ. Oviedo, Avda. Calvo Sotelo
     s/n, ES-33007 Oviedo, Spain
    \label{OVIEDO}}
\titlefoot{Department of Physics, University of Oxford,
     Keble Road, Oxford OX1 3RH, UK
    \label{OXFORD}}
\titlefoot{Dipartimento di Fisica, Universit\`a di Padova and
     INFN, Via Marzolo 8, IT-35131 Padua, Italy
    \label{PADOVA}}
\titlefoot{Rutherford Appleton Laboratory, Chilton, Didcot
     OX11 OQX, UK
    \label{RAL}}
\titlefoot{Dipartimento di Fisica, Universit\`a di Roma II and
     INFN, Tor Vergata, IT-00173 Rome, Italy
    \label{ROMA2}}
\titlefoot{Dipartimento di Fisica, Universit\`a di Roma III and
     INFN, Via della Vasca Navale 84, IT-00146 Rome, Italy
    \label{ROMA3}}
\titlefoot{DAPNIA/Service de Physique des Particules,
     CEA-Saclay, FR-91191 Gif-sur-Yvette Cedex, France
    \label{SACLAY}}
\titlefoot{Instituto de Fisica de Cantabria (CSIC-UC), Avda.
     los Castros s/n, ES-39006 Santander, Spain
    \label{SANTANDER}}
\titlefoot{Inst. for High Energy Physics, Serpukov
     P.O. Box 35, Protvino, (Moscow Region), Russian Federation
    \label{SERPUKHOV}}
\titlefoot{J. Stefan Institute, Jamova 39, SI-1000 Ljubljana, Slovenia
    \label{SLOVENIJA1}}
\titlefoot{Laboratory for Astroparticle Physics,
     University of Nova Gorica, Kostanjeviska 16a, SI-5000 Nova Gorica, Slovenia
    \label{SLOVENIJA2}}
\titlefoot{Department of Physics, University of Ljubljana,
     SI-1000 Ljubljana, Slovenia
    \label{SLOVENIJA3}}
\titlefoot{Fysikum, Stockholm University,
     Box 6730, SE-113 85 Stockholm, Sweden
    \label{STOCKHOLM}}
\titlefoot{Dipartimento di Fisica Sperimentale, Universit\`a di
     Torino and INFN, Via P. Giuria 1, IT-10125 Turin, Italy
    \label{TORINO}}
\titlefoot{INFN,Sezione di Torino and Dipartimento di Fisica Teorica,
     Universit\`a di Torino, Via Giuria 1,
     IT-10125 Turin, Italy
    \label{TORINOTH}}
\titlefoot{Dipartimento di Fisica, Universit\`a di Trieste and
     INFN, Via A. Valerio 2, IT-34127 Trieste, Italy
    \label{TRIESTE}}
\titlefoot{Istituto di Fisica, Universit\`a di Udine and INFN,
     IT-33100 Udine, Italy
    \label{UDINE}}
\titlefoot{Univ. Federal do Rio de Janeiro, C.P. 68528
     Cidade Univ., Ilha do Fund\~ao
     BR-21945-970 Rio de Janeiro, Brazil
    \label{UFRJ}}
\titlefoot{Department of Radiation Sciences, University of
     Uppsala, P.O. Box 535, SE-751 21 Uppsala, Sweden
    \label{UPPSALA}}
\titlefoot{IFIC, Valencia-CSIC, and D.F.A.M.N., U. de Valencia,
     Avda. Dr. Moliner 50, ES-46100 Burjassot (Valencia), Spain
    \label{VALENCIA}}
\titlefoot{Institut f\"ur Hochenergiephysik, \"Osterr. Akad.
     d. Wissensch., Nikolsdorfergasse 18, AT-1050 Vienna, Austria
    \label{VIENNA}}
\titlefoot{Inst. Nuclear Studies and University of Warsaw, Ul.
     Hoza 69, PL-00681 Warsaw, Poland
    \label{WARSZAWA}}
\titlefoot{Now at University of Warwick, Coventry CV4 7AL, UK
    \label{WARWICK}}
\titlefoot{Fachbereich Physik, University of Wuppertal, Postfach
     100 127, DE-42097 Wuppertal, Germany \\
\noindent
{$^\dagger$~deceased}
    \label{WUPPERTAL}}
\addtolength{\textheight}{-10mm}
\addtolength{\footskip}{5mm}
\clearpage

\headsep 30.0pt
\end{titlepage}

%
\pagenumbering{arabic}                              
\setcounter{footnote}{0}                            %
\large
\newcommand{\approxgt}{\raisebox{-.5ex}{\stackrel{>}{\sim}}}
\newcommand{\approxlt}{\raisebox{-.5ex}{\stackrel{<}{\sim}}}
\newcommand{\mean}[1]{{\left\langle #1 \right\rangle}}
\newcommand{\abs}[1]{{\left\|#1\right\|}}
\newcommand{\s}{{\sim}}
\newcommand{\ra}{{\rightarrow}}
\newcommand{\lra}{{\leftrightarrow}}
\newcommand{\longra}{{\longrightarrow}}
\newcommand{\mstrut}{\rule{0cm}{5ex}}
\newcommand{\ov}[1]{\overline{#1}}
\newcommand{\tc}[1]{{\mbox{\tiny #1}}}
\newcommand{\Real}{{\mathcal{R}e}}
\newcommand{\Imag}{{\mathcal{I}m}}


\newcommand{\bmW}{{\mathbf{W}}}
\newcommand{\bmF}{{\mathbf{F}}}
\newcommand{\bmX}{{\mathbf{X}}}
\newcommand{\bmI}{{\mathbf{I}}}
\newcommand{\bmalpha}{{\boldmath{\alpha}}}
\newcommand{\bmtheta}{{\mathbf{\theta}}}
\newcommand{\bmsigma}{{\mathbf{\sigma}}}
\newcommand{\bmlambda}{{\mathbf{\lambda}}}
\newcommand{\bmtau}{{\mathbf{\tau}}}
\newcommand{\bmdag}{{\mathbf{\dagger}}}
\newcommand{\Delx}{{\Delta x}}
\newcommand{\Dely}{{\Delta y}}
\newcommand{\rphi}{{r-\phi}}
\newcommand{\sz}{{s-z}}
\newcommand{\xy}{{x-y}}
\newcommand{\zo}{{z_0}}
\newcommand{\modzo}{{\left| z_0 \right|}}
\newcommand{\meanzo}{{\left\langle z_0 \right\rangle}}
\newcommand{\bphi}{{\bar\phi}}
\newcommand{\bthe}{{\bar\theta}}

\newcommand{\chisq}{{{\chi}^{2}}}
\newcommand{\chindof}{{\chisq / \mathrm{dof}}}

\newcommand{\grpsuth}{{\mathrm{SU}(3)}}
\newcommand{\grpsutw}{{\mathrm{SU}(2)}}
\newcommand{\grpuone}{{\mathrm{U}(1)}}

\newcommand{\qq}{\mathrm{q{\bar q}}}
\newcommand{\ppbar}{{p{\bar p}}}
\newcommand{\qqgam}{\mathrm{q{\bar q}(\gamma)}}
\newcommand{\ff}{{f{\bar f}}}
\newcommand{\lplm}{{\elll^{+}\elll^{-}}}
\newcommand{\bb}{{b{\bar b}}}
\newcommand{\cc}{{c{\bar c}}}
\newcommand{\ee}{{e^{+}e^{-}}}
\newcommand{\mumu}{{\mu^{+}\mu^{-}}}
\newcommand{\tautau}{{\tau^{+}\tau^{-}}}
\newcommand{\eemm}{{\ee \ra \mumu}}
\newcommand{\eemmg}{{\ee \ra \mumu \gamma}}
\newcommand{\eett}{{\ee \ra \tautau}}
\newcommand{\eeee}{{\ee \ra \ee}}
\newcommand{\eell}{{\ee \ra \lplm}}
\newcommand{\nl}{{\nu_{\ell}}}
\newcommand{\nbl}{{{\overline{\nu}}_{\ell}}}
\newcommand{\nmu}{{\nu_{\mu}}}
\newcommand{\nbmu}{{{\overline{\nu}}_{\mu}}}
\newcommand{\nel}{{\nu_{e}}}
\newcommand{\nbel}{{{\overline{\nu}}_{e}}}
\newcommand{\ntau}{{\nu_{\tau}}}
\newcommand{\nbtau}{{{\overline{\nu}}_{\tau}}}
\newcommand{\ppos}{{\pi^{+}}}
\newcommand{\pneg}{{\pi^{-}}}
\newcommand{\pzer}{{\pi^{0}}}
\newcommand{\ppm}{{\pi^{\pm }}}
\newcommand{\Zo}{{\mathrm{Z}^{0}}}
\newcommand{\Zostar}{{\mathrm{Z}^{0\ast}}}
\newcommand{\Wpm}{{\mathrm{W}^\pm}}
\newcommand{\Wp}{{\mathrm{W}^{+}}}
\newcommand{\Wm}{{\mathrm{W}^{-}}}
\newcommand{\W}{{\mathrm{W}}}
\newcommand{\Z}{{\mathrm{Z}}}
\newcommand{\ZZ}{{\mathrm{ZZ}}}
\newcommand{\Zee}{{\mathrm{Z \ee}}}
\newcommand{\Wen}{{\mathrm{W e \nel}}}
\newcommand{\Zgam}{{\mathrm{Z/\gamma}}}
\newcommand{\Ho}{{\mathrm{H}}}
\newcommand{\WW}{{\mathrm{ W^{+} W^{-} }}}
\newcommand{\epem}{{\mathrm{ e^{+} e^{-} }}}
\newcommand{\eeqq}{{\epem \ra \qqgam}}

\newcommand{\qqb}{\mathrm{q{\bar q'}}}
\newcommand{\qbq}{\mathrm{{\bar Q}Q'}}
\newcommand{\len}{{\ell \nbl}}
\newcommand{\mn}{{\mu \nbmu}}
\newcommand{\en}{{e \nbel}}
\newcommand{\tn}{{\tau \nbtau}}
\newcommand{\nmn}{{{\overline{\mu}} \nmu}}
\newcommand{\nen}{{{\overline{e}} \nel}}
\newcommand{\ntn}{{{\overline{\tau}} \ntau}}

\newcommand{\qqqq}{{{\mathrm q_1\bar{q_2}q_3\bar{q_4}}}}
\newcommand{\lnqq}{{\mathrm{ \len q_1 \bar{q_2}}}}
\newcommand{\enqq}{{\mathrm{ \en  \qqb}}}
\newcommand{\mnqq}{{\mathrm{ \mn  \qqb}}}
\newcommand{\tnqq}{{\mathrm{ \tn  \qqb}}}
\newcommand{\lnln}{{\ell \nbl {\overline{\ell}} \nl}}
\newcommand{\enen}{{\en \nen }}
\newcommand{\mnmn}{{\mn \nmn}}
\newcommand{\tntn}{{\tn \ntn}}
\newcommand{\enmn}{{\en \nmn}}
\newcommand{\entn}{{\en \ntn}}
\newcommand{\mntn}{{\mn \ntn}}

\newcommand{\mw}{{\mathrm{M_W}}}
\newcommand{\mwone}{{\mathrm{m_{\W 1}}}}
\newcommand{\mwtwo}{{\mathrm{m_{\W 2}}}}
\newcommand{\smw}{{\mathrm{m_W}}}
\newcommand{\mwav}{{\mathrm{\bar{m}_W}}}

\newcommand{\mz}{{\mathrm{M_Z}}}
\newcommand{\MZrp}[1]{{\mathrm{M_{\mbox{\tiny{Z}}}\!\!^{#1}}}}
\newcommand{\gz}{{\Gamma_{\mathrm{Z}}}}
\newcommand{\tz}{{\tau_{\mathrm{Z}}}}

\newcommand{\gw}{{\Gamma_{\mathrm{W}}}}
\newcommand{\mh}{{\mathrm{M_H}}}
\newcommand{\mt}{{\mathrm{m_t}}}
\newcommand{\ssqtw}{{\sin^{2}\!\theta_{\mathrm{W}}}}
\newcommand{\csqtw}{{\cos^{2}\!\theta_{\mathrm{W}}}}
\newcommand{\stw}{{\sin\theta_{\mathrm{W}}}}
\newcommand{\ctw}{{\cos\theta_{\mathrm{W}}}}
\newcommand{\ssqtwef}{{\sin}^{2}\theta_{\mathrm{W}}^{\mathrm{eff}}}
\newcommand{\ssqtfef}{{{\sin}^{2}\theta^{\mathrm{eff}}_{f}}}
\newcommand{\ssqtlef}{{{\sin}^{2}\theta^{\mathrm{eff}}_{l}}}
\newcommand{\csqtwef}{{{\cos}^{2}\theta_{\mathrm{W}}^{\mathrm{eff}}}}
\newcommand{\stwef}{\sin\theta_{\mathrm{W}}^{\mathrm{eff}}}
\newcommand{\ctwef}{\cos\theta_{\mathrm{W}}^{\mathrm{eff}}}
\newcommand{\gv}{{g_{\mbox{\tiny V}}}}
\newcommand{\ga}{{g_{\mbox{\tiny A}}}}
\newcommand{\gvel}{{g_{\mbox{\tiny V}}^{e}}}
\newcommand{\gael}{{g_{\mbox{\tiny A}}^{e}}}
\newcommand{\gvmu}{{g_{\mbox{\tiny V}}^{\mu}}}
\newcommand{\gamu}{{g_{\mbox{\tiny A}}^{\mu}}}
\newcommand{\gvf}{{g_{\mbox{\tiny V}}^{f}}}
\newcommand{\gaf}{{g_{\mbox{\tiny A}}^{f}}}
\newcommand{\gvl}{{g_{\mbox{\tiny V}}^{l}}}
\newcommand{\gal}{{g_{\mbox{\tiny A}}^{l}}}
\newcommand{\ghvf}{{\hat{g}_{\mbox{\tiny V}}^{f}}}
\newcommand{\ghaf}{{\hat{g}_{\mbox{\tiny A}}^{f}}}
\newcommand{\gvh}{{\hat{g}_{\mbox{\tiny V}}}}
\newcommand{\gah}{{\hat{g}_{\mbox{\tiny A}}}}
\newcommand{\thw}{{\theta_{\mbox{\mathrm{W}}}}}
\newcommand{\GF}{{G_{\mbox{\tiny F}}}}
\newcommand{\Vub}{{\mathrm{V}_{ub}}}
\newcommand{\Vcb}{{\mathrm{V}_{cb}=}}

\newcommand{\mgen}{{\mathrm{M_{gen}}}}

\newcommand{\ordalph}{{\mathcal{O}(\alpha)}}
\newcommand{\ordalsq}{{\mathcal{O}(\alpha^{2})}}
\newcommand{\ordalcb}{{\mathcal{O}(\alpha^{3})}}

\newcommand{\sigtot}{{\sigma_{\mbox{\tiny TOT}}}}
\newcommand{\sigf}{{\sigma_{\mbox{\tiny F}}}}
\newcommand{\sigb}{{\sigma_{\mbox{\tiny B}}}}
\newcommand{\dsigf}{{\delta\sigma_{\mbox{\tiny F}}}}
\newcommand{\dsigb}{{\delta\sigma_{\mbox{\tiny B}}}}
\newcommand{\dsfbi}{{\delta\sigma_{\mbox{\tiny fb}}^{\mbox{\tiny int}}}}
\newcommand{\AFB}{{A_{\mbox{\tiny FB}}}}
\newcommand{\Afbmm}{{A_{\mbox{\tiny FB}}^{\mbox{\tiny $\mu\mu$}}}}
\newcommand{\Afbtt}{{A_{\mbox{\tiny FB}}^{\mbox{\tiny $\tau\tau$}}}}
\newcommand{\APOL}{{A_{\mbox{\tiny POL}}}}
\newcommand{\pwff}{{\Gamma_{ff}}}
\newcommand{\pwee}{{\Gamma_{ee}}}
\newcommand{\pwmm}{{\Gamma_{\mu\mu}}}
\newcommand{\Af}{{\mathcal{A}_{f}}}
\newcommand{\Ael}{{\mathcal{A}_{e}}}
\newcommand{\Amu}{{\mathcal{A}_{\mu}}}
\newcommand{\dafbint}{{\delta A_{\mbox{\tiny FB}}^{\mbox{\tiny int}}}}
\newcommand{\vcs}{{\left| V_{cs} \right|} }

\newcommand{\Ephot}{{E_{\gamma}}}
\newcommand{\Ebeam}{{E_{\mbox{\tiny BEAM}}}}
\newcommand{\sqs}{{\protect\sqrt{s}}}
\newcommand{\sprime}{{\protect\sqrt{s^{\prime}}}}
\newcommand{\pT}{{\mathrm{p_T}}}
\newcommand{\mmu}{{m_{\mu}}}
\newcommand{\mb}{{m_{b}}}
\newcommand{\thacop}{{\theta_{\mbox{\tiny acop}}}}
\newcommand{\thacol}{{\theta_{\mbox{\tiny acol}}}}
\newcommand{\prad}{{p_{\mbox{\tiny rad}}}}
\newcommand{\lambdabar}{{\lambda \! \! \! \! {\raisebox{+.5ex}{$-$}}}}

\newcommand{\rprg}{{r_{\mbox{\tiny p}}}}
\newcommand{\zprg}{{z_{\mbox{\tiny p}}}}
\newcommand{\thprg}{{\theta_{\mbox{\tiny p}}}}
\newcommand{\phprg}{{\phi_{\mbox{\tiny p}}}}

\newcommand{\mrad}{{\mathrm{mrad}}}
\newcommand{\rad}{{\mathrm{rad}}}
\newcommand{\dgr}{{^\circ}}
\newcommand{\TeV}{{\mathrm{TeV}}}
\newcommand{\GeV}{{\mathrm{GeV}}}
\newcommand{\GeVm}{{\mathrm{GeV/c{^2}}}}
\newcommand{\GeVp}{{\mathrm{GeV/c}}}
\newcommand{\MeV}{{\mathrm{MeV}}}
\newcommand{\MeVp}{{\mathrm{MeV/c}}}
\newcommand{\MeVm}{{\mathrm{MeV/c{^2}}}}
\newcommand{\KeV}{{\mathrm{KeV}}}
\newcommand{\eV}{{\mathrm{eV}}}
\newcommand{\um}{{\mathrm{m}}}
\newcommand{\umm}{{\mathrm{mm}}}
\newcommand{\uum}{{\mu{\mathrm m}}}
\newcommand{\ucm}{{\mathrm{cm}}}
\newcommand{\ufm}{{\mathrm{fm}}}
\newcommand{\umicrom}{{\mathrm{\mu m}}}
\newcommand{\us}{{\mathrm{s}}}
\newcommand{\ums}{{\mathrm{ms}}}
\newcommand{\uus}{{\mu{\mathrm{s}}}}
\newcommand{\uns}{{\mathrm{ns}}}
\newcommand{\ups}{{\mathrm{ps}}}
\newcommand{\uub}{{\mu{\mathrm{b}}}}
\newcommand{\unb}{{\mathrm{nb}}}
\newcommand{\upb}{{\mathrm{pb}}}
\newcommand{\ipb}{{\mathrm{pb^{-1}}}}
\newcommand{\ifb}{{\mathrm{fb^{-1}}}}
\newcommand{\inb}{{\mathrm{nb^{-1}}}}

\newcommand{\std}{{\mathrm{std}}}
\newcommand{\cone}{{\mathrm{cone}}}
\newcommand{\cut}{{\mathrm{cut}}}
\newcommand{\Rcone}{{\mathrm{R_{cone}}}}
\newcommand{\Pcut}{{\mathrm{p_{cut}}}}
\newcommand{\dmwp}{{\Delta \mathrm{M_W}(\std,\mathrm{p_{cut}})}}
\newcommand{\dmwr}{{\Delta \mathrm{M_W}(\std,\mathrm{R_{cone}})}}
\newcommand{\dmwpct}{{\Delta \mathrm{M_W}(\std,\mathrm{p_{cut}}=2 \, \mathrm{GeV/c})}}
\newcommand{\dmwrcn}{{\Delta \mathrm{M_W}(\std,\mathrm{R_{cone}}=0.5 \, \mathrm{rad})}}
\newcommand{\mwp}{{\mathrm{M_W}^{\mathrm{p_{cut}}}}}
\newcommand{\mwr}{{\mathrm{M_W}^{\mathrm{R_{cone}}}}}
\newcommand{\mwpct}{{\mathrm{M_W}^{\mathrm{p_{cut}}=2 \, \mathrm{GeV/c}}}}
\newcommand{\mwrcn}{{\mathrm{M_W}^{\mathrm{R_{cone}}=0.5 \, \mathrm{rad}}}}
\newcommand{\eg}{\mbox{\itshape e.g.}}
\newcommand{\ie}{\mbox{\itshape i.e.}}
\newcommand{\etal}{{\slshape et al\/}\ }
\newcommand{\etc}{\mbox{\itshape etc}}
\newcommand{\cf}{\mbox{\itshape cf.}}
\newcommand{\ffp}{\mbox{\itshape ff}}
\newcommand{\cm}{\mbox{c.m.}}
\newcommand{\ALEPH}{\mbox{ALEPH}}
\newcommand{\DELPHI}{\mbox{DELPHI}}
\newcommand{\OPAL}{\mbox{OPAL}}
\newcommand{\LTHREE}{\mbox{L3}}
\newcommand{\CERN}{\mbox{CERN}}
\newcommand{\LEP}{\mbox{LEP}}
\newcommand{\LEPONE}{\mbox{LEP-1}}
\newcommand{\LEPTWO}{\mbox{LEP-2}}
\newcommand{\CDF}{\mbox{CDF}}
\newcommand{\DO}{\mbox{D0}}
\newcommand{\SLD}{\mbox{SLD}}
\newcommand{\CLEO}{\mbox{CLEO}}
\newcommand{\UAONE}{\mbox{UA1}}
\newcommand{\UATWO}{\mbox{UA2}}
\newcommand{\TEVATRON}{\mbox{TEVATRON}}
\newcommand{\LHC}{\mbox{LHC}}
\newcommand{\KORALZ}{\mbox{\ttfamily KORALZ}}
\newcommand{\KORALW}{\mbox{\ttfamily KORALW}}
\newcommand{\ZFITTER}{\mbox{\ttfamily ZFITTER}}
\newcommand{\GENTLE}{\mbox{\ttfamily GENTLE}}
\newcommand{\DELANA}{\mbox{\ttfamily DELANA}}
\newcommand{\DELSIM}{\mbox{\ttfamily DELSIM}}
\newcommand{\DYMU}{\mbox{\ttfamily DYMU3}}
\newcommand{\TANAGRA}{\mbox{\ttfamily TANAGRA}}
\newcommand{\ZEBRA}{\mbox{\ttfamily ZEBRA}}
\newcommand{\PAW}{\mbox{\ttfamily PAW}}
\newcommand{\WWANA}{\mbox{\ttfamily WWANA}}
\newcommand{\FASTSIM}{\mbox{\ttfamily FASTSIM}}
\newcommand{\PYTHIA}{\mbox{\ttfamily PYTHIA}}
\newcommand{\JETSET}{\mbox{\ttfamily JETSET}}
\newcommand{\TWOGAM}{\mbox{\ttfamily TWOGAM}}
\newcommand{\LUBOEI}{\mbox{\ttfamily LUBOEI}}
\newcommand{\SKI}{\mbox{\ttfamily SK-I}}
\newcommand{\SKII}{\mbox{\ttfamily SK-II}}
\newcommand{\ARIADNE}{\mbox{\ttfamily ARIADNE}}
\newcommand{\ARII}{\mbox{\ttfamily AR-II}}
\newcommand{\VNI}{\mbox{\ttfamily VNI}}
\newcommand{\HERWIG}{\mbox{\ttfamily HERWIG}}
\newcommand{\EXCALIBUR}{\mbox{\ttfamily EXCALIBUR}}
\newcommand{\WPHACT}{\mbox{\ttfamily WPHACT}}
\newcommand{\RACOONWW}{\mbox{\ttfamily RACOONWW}}
\newcommand{\YFSWW}{\mbox{\ttfamily YFSWW}}
\newcommand{\QEDPS}{\mbox{\ttfamily QEDPS}}
\newcommand{\CCTHREE}{\mbox{\ttfamily CC03}}
\newcommand{\NCEIGHT}{\mbox{\ttfamily NC08}}
\newcommand{\LUCLUS}{\mbox{\ttfamily LUCLUS}}
\newcommand{\JADE}{\mbox{\ttfamily JADE}}
\newcommand{\DURHAM}{\mbox{\ttfamily DURHAM}}
\newcommand{\CAMBRIDGE}{\mbox{\ttfamily CAMBRIDGE}}
\newcommand{\DICLUS}{\mbox{\ttfamily DICLUS}}
\newcommand{\CONE}{\mbox{\ttfamily CONE}}
\newcommand{\PUFITC}{\mbox{\ttfamily PUFITC+}}
\newcommand{\PHDST}{\mbox{\ttfamily PHDST}}
\newcommand{\SKELANA}{\mbox{\ttfamily SKELANA}}
\newcommand{\DAFNE}{\mbox{\ttfamily DAFNE}}
\newcommand{\CERNLIB}{\mbox{\ttfamily CERNLIB}}
\newcommand{\MINUIT}{\mbox{\ttfamily MINUIT}}
\newcommand{\REMCLU}{\mbox{\ttfamily REMCLU}}
\newcommand{\DST}{\mbox{\ttfamily DST}}
\newcommand{\XSDST}{\mbox{\ttfamily XShortDST}}
\newcommand{\FDST}{\mbox{\ttfamily FullDST}}

\newcommand{\spot}{\mbox{$\bullet \;$}}

\newcommand{\eps}{{\epsilon}}
\newcommand{\erreps}{{\sigma_{\epsilon}}}
\newcommand{\errepsp}{{\sigma_{\epsilon}^{+}}}
\newcommand{\errepsm}{{\sigma_{\epsilon}^{-}}}
\newcommand{\Lmb}{{\Lambda}}
\newcommand{\lmb}{{\lambda}}
\newcommand{\Lmbsq}{{\Lambda^{2}}}
\newcommand{\Lmbisq}{{1/\Lambda^{2}}}
\newcommand{\Lmbpm}{{\Lambda^{\pm}}}
\newcommand{\Lmbp}{{\Lambda^{+}}}
\newcommand{\Lmbm}{{\Lambda^{-}}}
\newcommand{\gc}{{g_{c}}}
\newcommand{\etaij}{{\eta_{ij}}}
\newcommand{\IJ}{{\mathrm{IJ}}}
\newcommand{\IJpm}[1]{{\mathrm{IJ}^{(#1)}}}

\newcommand{\gev}{{\mathrm{GeV}}}
\newcommand{\wpm}{{\mathrm{W}^\pm}}
\newcommand{\wm}{{\mathrm{W}^{-}}}
\newcommand{\w}{{\mathrm{W}}}
\newcommand{\z}{{\mathrm{Z}}}
\newcommand{\ww}{{\mathrm{ W^{+} W^{-} }}}
\newcommand{\tev}{{\mathrm{TeV}}}
\newcommand{\lep}{\mbox{LEP}}
\newcommand{\lepone}{\mbox{LEP-1}}
\newcommand{\leptwo}{\mbox{LEP-2}}
\newcommand{\delphi}{\mbox{DELPHI}}
\newcommand{\excalibur}{\mbox{\ttfamily EXCALIBUR}}
\newcommand{\wwana}{\mbox{\ttfamily WWANA}}
\newcommand{\fastsim}{\mbox{\ttfamily FASTSIM}}
\newcommand{\phdst}{\mbox{\ttfamily PHDST}}
\newcommand{\skelana}{\mbox{\ttfamily SKELANA}}
\newcommand{\dafne}{\mbox{\ttfamily DAFNE}}
\newcommand{\delana}{\mbox{\ttfamily DELANA}}
\newcommand{\zebra}{\mbox{\ttfamily ZEBRA}}
\newcommand{\delsim}{\mbox{\ttfamily DELSIM}}
\newcommand{\pythia}{\mbox{\ttfamily PYTHIA}}
\newcommand{\jetset}{\mbox{\ttfamily JETSET}}
\newcommand{\ariadne}{\mbox{\ttfamily ARIADNE}}
\newcommand{\cdf}{\mbox{CDF}}
\newcommand{\gentle}{\mbox{\ttfamily GENTLE}}

\def\Journal#1#2#3#4{{#1}{\bf #2}(#4) #3}
\def\PLB{{ Phys. Lett.}  \bf B}
\def\EUR{{ Eur. Phys. J.} \bf C}
\def\PRL{{ Phys. Rev. Lett.} }
\def\NIMA{{Nucl. Instr. and Meth.} \bf A}
\def\PRD{{ Phys. Rev.} \bf D}
\def\ZPC{{ Zeit. Phys.} \bf C}
\def\CPC{{Comput. Phys. Commun.} }

\def\bb{{b\bar{b}}}
\def\cc{{c\bar{c}}}
\def\ll{{l\bar{l}}}
\def\lbl{{l\bar{l}}}

\newcommand{\kos}{\ifmmode {{\mathrm K}^{0}_{S} } \else
${\mathrm K}^{0}_{S}$ \fi}
\newcommand{\kpm}{\ifmmode {{\mathrm K}^{\pm}} \else
${\mathrm K}^{\pm}$\fi}
\newcommand{\ko}{\ifmmode {{\mathrm K}^{0}} \else
${\mathrm K}^{0}$ \fi}
\def\Lam{$\Lambda$ }

\def\Rmu{$R_3(\mu)/R_3(had)$\ }
\def\Re{$R_3(e)/R_3(had)$\ }
\def\Rmue{$R_3(\mu +e)/R_3(had)$\ }

\def\ee{$e\sp{+}e\sp{-}$}
\newcommand{\Wfj}{WW $\rightarrow 4-{\mathrm{jets}}$ events}
\newcommand{\Wtj}{WW $\rightarrow 2-{\mathrm{jets}} \, \ell \bar{\nu}$ events}

\def\rf#1{\ref{fig:#1}}
\def\micap#1{\caption{\sl #1}}
\def\bfg#1{\begin{figure}[#1] \begin{center}}
\def\efg#1{\label{fig:#1} \end{center} \end{figure}}
\newcommand{\KK}{\mbox{\ttfamily KK2f}}

\newcommand{\chisqfc}{{\chi^2_{4C}}}
\newcommand{\pkin}{{\bar p_j}}
\newcommand{\WWqqqq}{{\WW~\ra~\qqqq}}
\newcommand{\WWffff}{{\WW~\ra~\ffff}}
\newcommand{\Zqqgam}{{\Z~\ra~\qqgam}}
\newcommand{\ZZqqqq}{{\ZZ~\ra~\qqqq}}


\section{Introduction}
\label{sec:intro}

The space-time development of a hadronic system is still poorly understood, and
models are necessary to transform a partonic system, governed by
perturbative QCD, to final state hadrons observed in the detectors.

WW events produced in $e^+e^-$ collisions at LEP-2
 constitute a unique laboratory
to study and test the evolution of such hadronic systems,
because of the clean environment and the well-defined initial energy
in the process. Of particular interest is the possibility to
study separately one single evolving hadronic system (one of the W
bosons decaying semi-leptonically, the other decaying hadronically), and
compare it with two hadronic systems evolving at the same time
(both W bosons decaying hadronically).

Interconnection effects between the products of the hadronic decays of
the two W bosons (in the same event) are expected since
the lifetime of the W bosons
($\tau_{\W} \simeq \hbar /\Gamma_{\W} \simeq 0.1$ fm/c)
is an order of magnitude smaller than the typical hadronization times.
These effects can happen at two levels:
\begin{itemize}
\item in the evolution of the parton shower,
between partons from different hadronic systems by exchanging coloured
gluons~\cite{ghz}
(this effect is called {\em Colour Reconnection} (CR) for historical reasons);
\item between the final state hadrons, due to quantum-mechanical interference,
mainly due to Bose-Einstein
Correlations (BEC) between identical bosons (e.g. pions with the same charge).
\end{itemize}
A detailed study by DELPHI of this second effect was recently
published~\cite{becww}.


The first effect, the possible presence of colour flow between the
two W hadronization systems, is the topic
studied in this paper. This effect is worthy of study in its own right and
for the possible effects induced on the W mass measurement in fully hadronic
events (see for instance \cite{lep2yr} for an introduction and
\cite{review} for an experimental review).

The effects at the perturbative level are expected to be small~\cite{lep2yr},
whereas they may be large at the hadronization level (many soft gluons sharing
the space-time) for which models have to be used to compare with the data.

The most tested model is the Sj\"ostrand-Khoze ``Type 1'' CR model
\SKI~\cite{sk1}. This model of CR is based on the Lund string
fragmentation phenomenology. The strings are considered as colour
flux tubes with some volume, and reconnection occurs when these
tubes overlap. The probability of reconnection in an event is
parameterised by the value $\kappa$, set globally by the user,
according to the space-time volume overlap of the two strings,
$V_{\mathrm{overlap}}\,$: \begin{equation}
\label{eq:sk1} 
{\cal P}_{\mathrm{reco}}(\kappa)=1-e^{-\kappa V_{\mathrm{overlap}}}\, .
\end{equation}
The parameter $\kappa$ was introduced in the \SKI\ model to allow a
variation of the percentage
of reconnected events and facilitate studies of sensitivity to the effect.
In this model only one string reconnection per event was allowed. The authors
of the model
propose the value of $\kappa=0.66$ to give similar amounts of reconnection
as other models of Colour Reconnection.
By comparing the data with the model predictions evaluated
at several $\kappa$ values, it is possible to determine the value of $\kappa$
most consistent
with the data and extract the corresponding reconnection probability.
Another model was proposed by the same authors, considering the colour flux tubes as
infinitely thin, which
allows for Colour Reconnection in the case the tubes cross each other and provided
the total string length is reduced (\SKII$^\prime$).
This last model was not tested.

Two further models are tested here, these are the models
implemented in \HERWIG~\cite{herwig} and \ARIADNE~\cite{ariadne}
Monte Carlo programs. In \HERWIG\ the partons are reconnected,
with a reconnection probability of 1/9, if the reconnection
results in a smaller total cluster mass. In \ARIADNE, which
implements an adapted version of the Gustafson-H\"akkinen
model~\cite{CR-GH}, the model used~\cite{AR-2} allows for
reconnections between partons originating in the same W boson, or
from different W bosons if they have an energy smaller than the
width of the W boson (this model will be referred as `AR-2').

Colour Reconnection has been previously investigated in DELPHI by comparing
inclusive distributions of charged particles, such as the
charged-particle multiplicity distribution
or the production of identified (heavy) particles, in fully hadronic WW events
and the distributions in semi-leptonic WW events.
The investigations did not show any effect as they were limited by statistical
and systematic errors and excluded only the most extreme models of CR
(see~\cite{delww}).

This article presents the results of the investigations of
Colour Reconnection effects in
hadronically decaying W pairs using two techniques.
The first, proposed by L3 in~\cite{L3du}, looks
at the particle flow between the jets in a 4-jet WW event. The second, proposed by DELPHI
in~\cite{JDH_MWnote}, takes into account the different sensitivity to Colour
Reconnection of several W mass estimators.
The first technique is more independent of the model and it
can provide comparisons based on data.
The second technique is more dependent on the model tested, but has
a much larger sensitivity to the models \SKI\ and \HERWIG.
Since the particle flow and W mass estimator methods were found to be largely
uncorrelated a combination of the results of these two methods is provided.

The paper is organised as follows. In the next section, the LEP
operation and the components of the DELPHI detector relevant to
the analyses are briefly described. In section 3 data and
simulation samples are explained. Then both of the analysis
methods discussed above are described and their results presented
in sections 4 and 5. The combination of the results is given in
section 6 and conclusions are drawn in the seventh and final
section.

\section{LEP Operation and Detector Description}
\label{sec:det}

At LEP-2, the second phase of the $\epem$ collider at CERN, the
accelerator was operated at
centre-of-mass energies above the threshold for double W boson production
from 1996 to 2000. In this period, the DELPHI experiment
collected about 12000 WW
events corresponding to a total integrated luminosity of 661 pb$^{-1}$.
About 46\% of the WW events are
WW $\rightarrow {\mathrm{q_1\bar{q}_2q_3\bar{q}_4}}$ events (fully hadronic),
and 44\% are
WW $\rightarrow {\mathrm{q_1\bar{q}_2\ell\bar{\nu}}}$, where $\ell$ is a
lepton (semi-leptonic).

The detailed description of the DELPHI detector and its performance is
provided in~\cite{deldet,perfo}.
A brief summary of the main characteristics of the detector important
for the analyses follows.

The tracking system of DELPHI consisted of a Time Projection Chamber (TPC), the main
tracking device of DELPHI, and was complemented by
a Vertex Detector (VD) closest to the beam pipe, the Inner and the Outer Detectors
in the barrel region, and two Forward Chambers in the end caps. It was embedded in a
1.2~T magnetic field, aligned parallel to the beam axis.

The electromagnetic calorimeter consisted of the
High density Projection Chamber (HPC) in the barrel region, the Forward
Electromagnetic Calorimeter (FEMC) and the Small angle Tile Calorimeter (STIC)
in the forward regions, complemented by detectors 
to tag the passage of electron-positron pairs from photons converted in
the regions between
the FEMC and the HPC. The total
depths of the calorimeters corresponded to about 18 radiation lengths.
The hadronic calorimeter was composed of instrumented iron
with a total depth along the shortest trajectory for a neutral
particle of 6 interaction lengths, and covered 98\% of the total
solid angle. Embedded in the hadronic calorimeter were two planes of
muon drift chambers to tag the passage of muons.
The whole detector was surrounded by a further double plane of staggered muon drift chambers.

For LEP-2, the DELPHI detector was upgraded as described
in the following.

Changes were made to some of the subdetectors, the trigger system~\cite{lep2trigger},
the run control and the algorithms used in the offline reconstruction
of tracks, which improved the performance compared to the earlier LEP-1 period.

The major changes were the extensions of the Vertex Detector (VD) and the Inner Detector (ID),
and the inclusion of the Very Forward Tracker
(VFT)~\cite{vft}, which increased the coverage of the
silicon tracker to polar angles with respect
to the $z$-axis\footnote{The \DELPHI\
coordinate system is a right-handed system with the $z$-axis collinear with
the incoming electron beam, and the $x$ axis pointing to the centre of the LEP
accelerator.} of $11\dgr < \theta < 169\dgr$.
To further improve the track reconstruction efficiency in the forward
regions of \DELPHI, the tracking algorithms and the
 alignment and calibration procedures were optimised for LEP-2.

Changes were also made to the electronics of the trigger and
timing system which improved the stability of the running during
data taking. The trigger conditions were optimised for LEP-2
running, to give high efficiency for 2- and 4-fermion processes in
the Standard Model and also to give sensitivity to events which
may have been signatures of new physics. In addition, improvements
were made to the operation of the detector during the LEP
operating states, to prepare the detector for data taking at the
very start of stable collisions of the $\epem $ beams, and to
respond to adverse background from LEP when it arose. These
changes led to an overall improvement
in the efficiency for collecting the delivered luminosity from
about $85\%$ in 1995, before the start of LEP-2,  to about $95\%$ at the
end in 2000.

During the operation of the \DELPHI\ detector in 2000 one of the 12 sectors of
the central tracking chamber, the TPC, failed. After
$1^{\mathrm{st}}$ September it was not possible to detect the tracks left
by charged particles inside the broken sector. The data affected corresponds
to around $1/4$ of the data collected in 2000. Nevertheless, the redundancy
of the tracking system of \DELPHI\ meant that tracks passing through the sector
could still be reconstructed from signals in any of the other tracking
detectors.
As a result, the track reconstruction efficiency was only slightly reduced in
the region covered by the broken sector, but the
track parameter resolutions 
were degraded compared with the data taken prior to the failure of this sector.

\section{Data and Simulation Samples}
\label{sec:datasim}

The analyses presented here use the data collected by DELPHI in the years
1997 to 2000, at centre-of-mass energies $\sqrt{s}$
between 183 and 209~GeV.
The data collected in the year 2000 with the TPC working in full,
with centre-of-mass energies from 200 to 208 GeV and a
integrated luminosity weighted average centre-of-mass energy of 206 GeV,
were analysed together.
Data acquired with the TPC with a broken sector, corresponding to a
integrated luminosity weighted average
centre-of-mass energy of 207 GeV, were analysed separately
and included in the results presented here.

The total integrated luminosity of the data sample
is 660.8 pb$^{-1}$, and the integrated luminosity weighted average centre-of-mass
energy of the data is 197.1~GeV.

To compare with the expected results from processes in the Standard Model
including or not including CR, Monte Carlo (MC) simulation was used
to generate events and simulate the response of the DELPHI detector.
These events were reconstructed and analysed with the same programs as
used for the real data.

The 4-fermion final states were generated with the code described in~\cite{wphdelphi},
based on \WPHACT~\cite{wphact},
for the WW signal (charged currents) and for the ZZ
background (neutral currents),
after which the events were fragmented with \PYTHIA~\cite{lund}
tuned to DELPHI data~\cite{tuning}.
The same WW events generated at 189, 200 and 206 GeV were also fragmented with
\PYTHIA\ implementing the \SKI\ model, with 100\% reconnection probability.
The systematic effects of fragmentation were studied using the above WW
samples and WW samples
generated with \WPHACT\ and fragmented with either \ARIADNE~\cite{ariadne} or
\HERWIG~\cite{herwig} at 183, 189, 200 and 206 GeV.
For systematic studies of Bose-Einstein Correlations (BEC), WW samples
generated with \WPHACT\ and fragmented with \PYTHIA\ implementing the BE$_{32}$
model~\cite{bec3} of BEC,
were used at all energies, except at 207 GeV.
The integrated luminosity of the simulated
samples was at least 10 times that of the data of the corresponding year, and the
majority corresponded to 100 times that of the data.

To test the consistency of the \SKI\ model and measure the $\kappa$ parameter,
large WW samples were generated in an early stage
of this work with \EXCALIBUR~\cite{excalibur} at 200 and 206 GeV,
keeping only the fully hadronic decays. These samples were then fragmented
with \PYTHIA. It was verified for smaller subsets that the results using these
large samples and the samples generated later with \WPHACT\ are compatible.

The ${\mathrm q}\bar{\mathrm q}(\gamma)$ background events were generated at
all energies with \KK~\cite{kk2f} and fragmented
with \PYTHIA. 
For systematic studies, similar \KK\ samples fragmented with
\ARIADNE~\cite{ariadne} 
were used
at 183, 189, 200 and 206 GeV.

These samples will be referred to as ``DELPHI samples''.

At 189 GeV, to compare with the other LEP experiments and with different
CR models, 6 samples generated with \KORALW~\cite{koralw} for the
4-fermion final states were also used.
These samples \footnote{produced by ALEPH after the LEP-W Physics
Workshop in Cetraro, Italy, October 2001} will be referred to as ``Cetraro samples''.
The events in the different samples have the final state quarks generated
with the same kinematics, and differ only
in the parton shower evolution and fragmentation.
Three samples were fragmented respectively with \PYTHIA, \ARIADNE\ and
\HERWIG\ (using the tuning of the ALEPH collaboration),
with no CR implementation.
Three other samples were fragmented in the same manner but now implementing
several CR models: the \SKI\ model with 100\% reconnection
probability, the AR-2 model, and the \HERWIG\ implementation of CR
with $1/9$ of reconnected events, respectively.


\section{The Particle Flow Method}

The first of the two analyses presented in this paper is based on the
so-called ``particle flow method''.
The particle flow algorithm is based on the selection of special event
topologies, in order to obtain
well defined regions between any two jets originating from
the same W (called the Inside-W region) or from
different Ws (called the Between-W region).
It is expected that Colour Reconnection decreases (increases)
particle production in the Inside-W (Between-W) region.
Hence, by studying the particle production in the inter-jet regions it is
possible to measure the effects of Colour Reconnection. However, this method
requires a selection of events with a suitable topology (see below) which has
a low efficiency (\raisebox{-.5ex}{$\stackrel{<}{\sim}$}25\%).

\subsection{Event and Particle Selection}

Events with both Ws decaying into $\mathrm{q_1 \bar q_2}$ are
characterised by high multiplicity, large visible energy, and the
tendency of the particles to be grouped in 4 jets. The background
is dominated by q$\bar{\mathrm q}(\gamma)$ events.

Charged particles were required to have momentum $p$ larger than 100~$\MeVp$
and below 1.5 times the beam energy,
a relative error on the momentum measurement
\mbox{$\Delta p/p<1$,} and
polar angle $\theta$ with respect to the beam axis between $20^\circ$ and 160$^\circ$.
To remove tracks from secondary interactions, the
distance of closest approach of the extrapolated track to the interaction point
was required to be less than
4~cm in the plane perpendicular to the beam axis
and less than 4/$\sin\theta$~cm along the beam axis, and the
reconstructed track length was required to be larger than 30~cm.

Clusters in the electromagnetic or hadronic calorimeters with energy
larger
than 0.5 GeV and polar angle in the interval $10^\circ <\theta< 170^\circ$,
not associated to charged particles, were considered as neutral particles.

\begin{figure}[H] 
\begin{center}
\mbox{\epsfxsize0.485\linewidth\epsffile{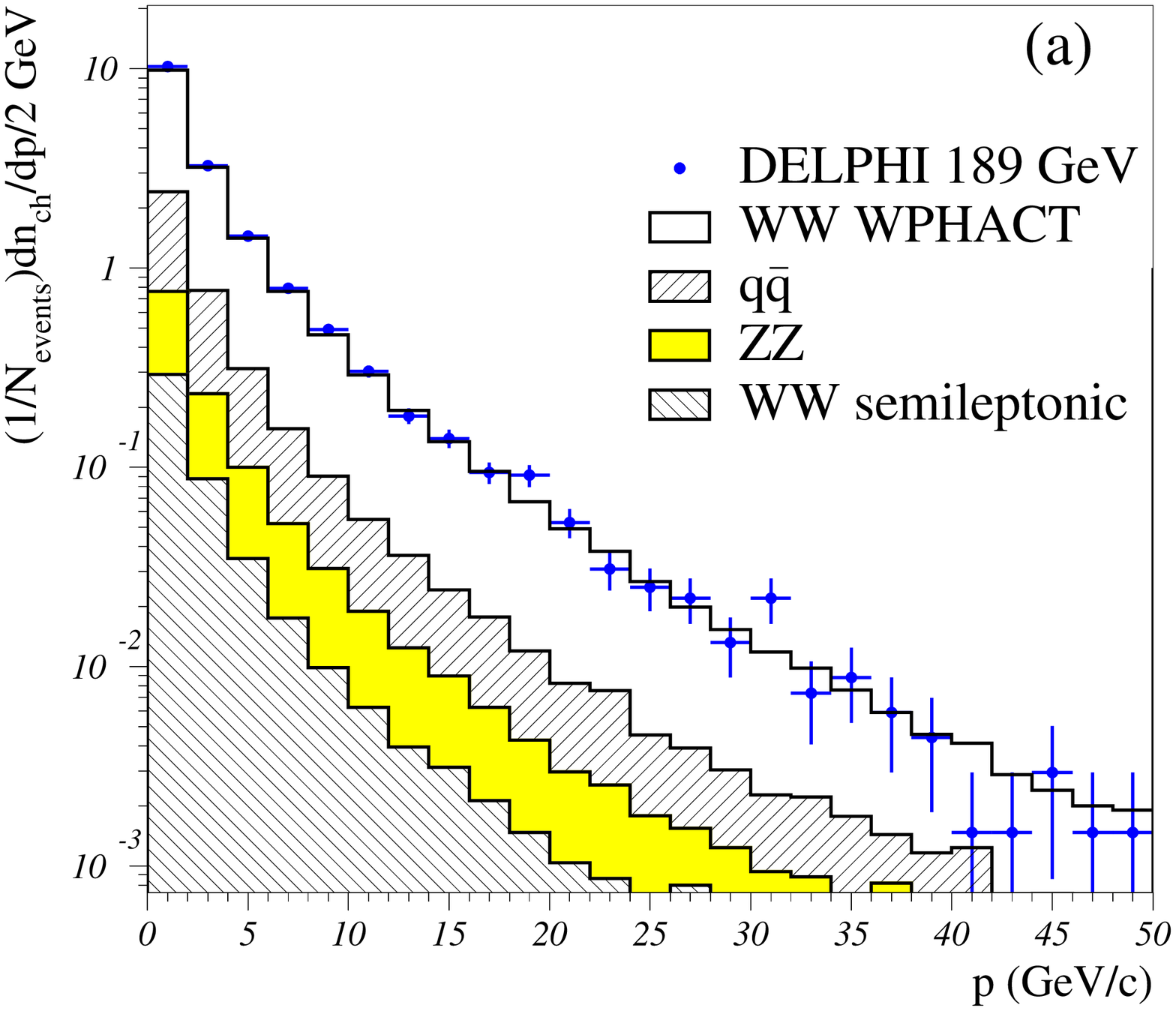}}
\mbox{\epsfxsize0.485\linewidth\epsffile{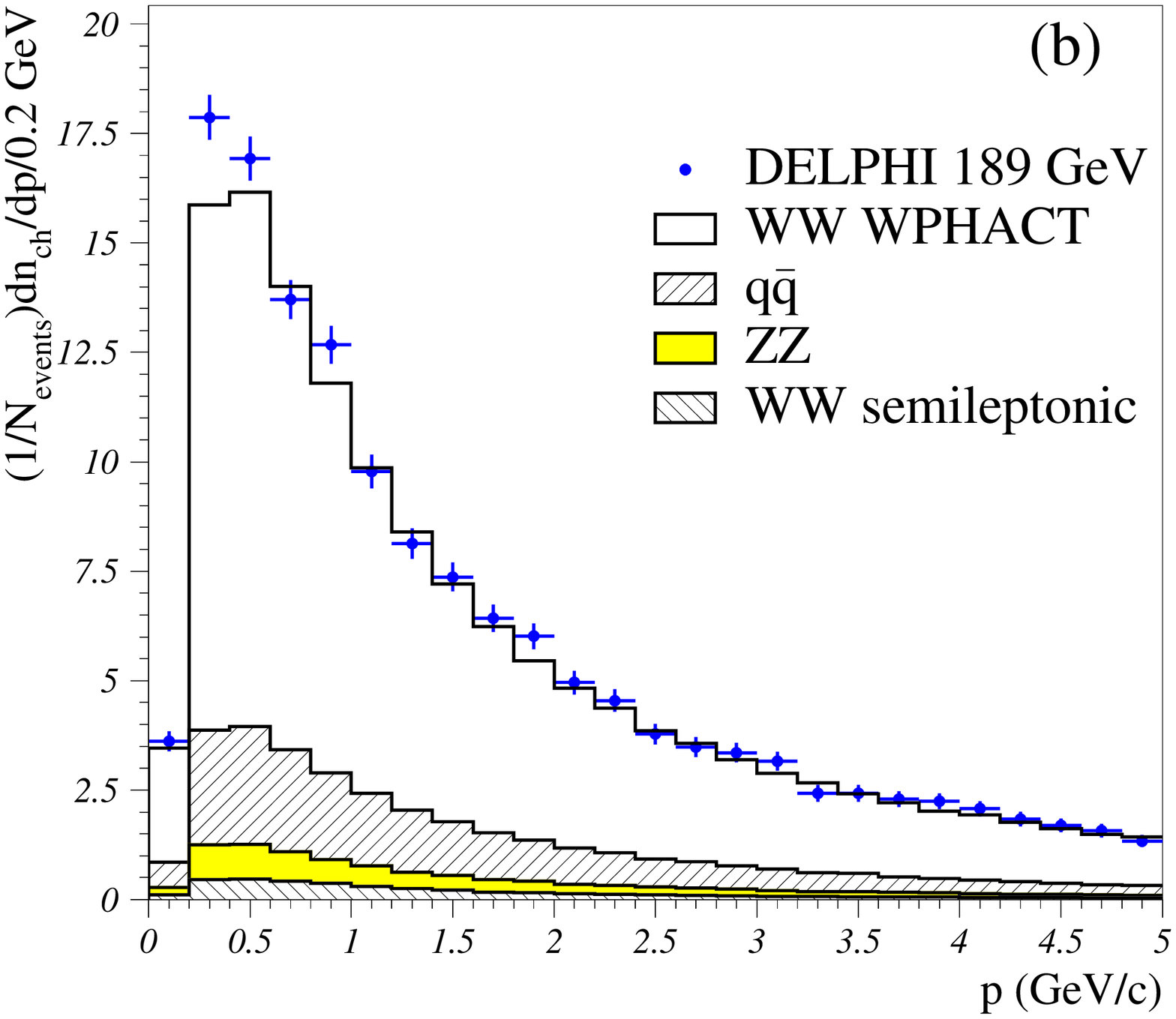}}
\caption[]{Momentum distribution for charged particles
(range 0-50~$\GeVp$ (a) and 0-5~$\GeVp$ (b)). Points represent the
data and the histograms represent the contributions from simulation for
the different processes (signal (white) and background contributions).
}\label{momenta}
\end{center}
\end{figure}

The events were pre-selected by requiring at least 12 charged particles, with
a sum of the modulus of the momentum transverse to the beam axis,
of charged and neutral
particles, above 20\%
of the centre-of-mass energy. These cuts reduced the
contributions from gamma-gamma processes and beam-gas interactions
to a negligible amount.
The momentum distribution of the charged particles for the
pre-selected events is shown
 in Figure~\ref{momenta}
and compared to the expected distribution from the simulation.
A good agreement between data and simulation is observed.

About half of the $\eeqq $ events at high-energy are associated
with an energetic photon emitted by one of the beam electrons or positrons
(radiative return events), thus reducing the energy available in the hadronic
system to the $\Z$ mass.
To remove these radiative return
events, the effective
centre-of-mass energy
\mbox{$\sqrt{s^\prime}$}, computed as described in \cite{sprime},
was required to be above 110~$\GeV$. It was verified that this cut does not affect
the signal from W pairs, but reduces significantly the contribution from
the q$\bar{\mathrm q}(\gamma)$ process.

\begin{figure}[hbt]
\begin{center}
\mbox{\epsfxsize10cm\epsffile{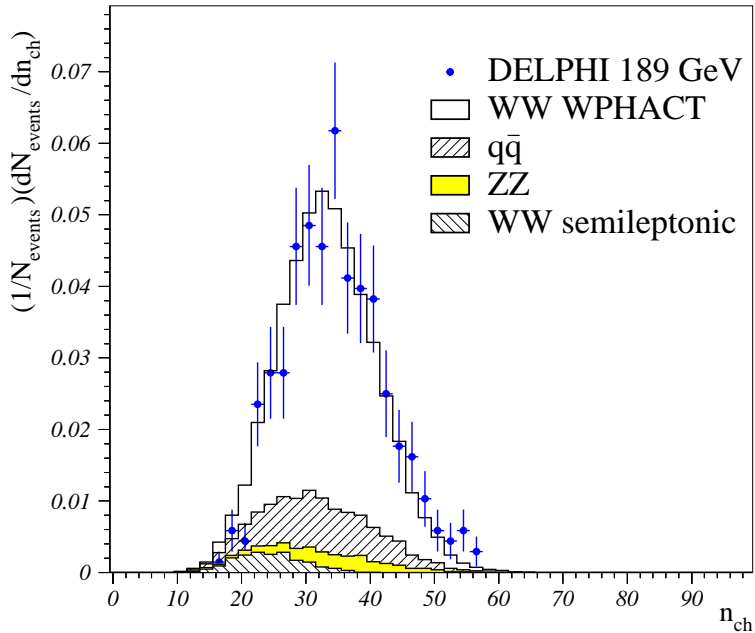}}
\caption[]{Uncorrected charged-particle multiplicity distribution at a
centre-of-mass energy of 189 GeV. Points represent the data and
the histograms represent the contribution from simulation
for the different processes.
}\label{mult189}
\end{center}
\end{figure}

In the WW fully hadronic decays four well separated energetic jets are
expected which balance the momentum of the event and have a total energy
near to the centre-of-mass energy.
The charged and neutral particles in the event were thus clustered 
using the 
DURHAM algorithm~\cite{durham},
for a separation value of 
$y_{\mathrm{cut}}=0.005$,
and the events were kept if there were 4 and only 4 jets and a multiplicity
(charged plus neutral) in each jet larger than 3.
The combination of these two cuts removed most of the
semi-leptonic WW
decays and the 2-jet and 3-jet events of the q$\bar{\mathrm q}(\gamma)$
background.
%
%
%
The charged-particle multiplicity distribution for the selected events at 189~GeV
is given in Figure~\ref{mult189}, with data points compared to the histogram from simulation
of signal and background processes.

For the study of the charged-particle flow between jets, the initial quark
configuration
should be well reconstructed with a good quark-jet
association. At 183 GeV and above, the produced W
bosons are significantly boosted. This produces smaller angles
in the laboratory frame of reference between the jets into which the W
decays, when compared to these angles at threshold (back-to-back).
Hence, this property
tends to reduce the ambiguity in the definition of the Between-W and
Inside-W regions.
The selection criteria were designed in order to minimize the situation of one jet
from one W boson appearing in the Inside-W region of the other W boson.

The selection criteria are based on the event topology, with
cuts in 4 of the 6 jet-jet angles. The
smallest and the second smallest jet-jet angle
should be below 100$^\circ$
and not adjacent (not have a common jet).
Two other jet-jet angles should be
between 100$^\circ$ and 140$^\circ$ and not adjacent (large angles).

In the case that there are two different combinations of jets satisfying
the above
criteria for the large angles, the combination with the highest sum of
large angles is chosen.
This selection increases the probability to have a correct pairing of jets
to the same W boson.

The integrated luminosity, the efficiency to select 4-jet WW
events and the purity of the selected data samples, estimated
using simulation, and the number of selected events are summarised
for each centre-of-mass energy in Table~\ref{tab:pureff}. The
numbers of expected events are also given separately for the
signal and the background processes, and were estimated using
simulation.
The efficiency to select the correct pairing of jets to the same W boson,
estimated with simulation as the fraction of WW events for which
the selected jets
1 and 2 (see later) correspond indeed to the same W boson, is given in the
last column of the Table.

\begin{table}
\begin{center}
\begin{tabular}{|c|c|c|c|c|c|c|c|c|c|c|}
\hline
$\sqrt{s}$
    &$\cal L$
            & Eff.& Pur.  & $N_{\mathrm{sel}}$
                                & MC tot.
                                        & WW 4j & $\mathrm{q\bar q}(\gamma)$
                                                       & ZZ  & W lep.
                                                                   & $\varepsilon_{\mathrm{PAIR}}$\\ \hline
\hline
183 &  52.7 & 22\% & 74\% & 127 & 114.2 &  84.4 & 22.3 & 0.7 & 7.0 & 69\% \\ \hline
189 & 157.6 & 21\% & 75\% & 340 & 341.4 & 255.9 & 56.8 & 2.4 &26.4 & 75\% \\ \hline
192 &  25.9 & 21\% & 75\% &  61 &  56.1 &  41.9 &  9.4 & 0.4 & 4.4 & 77\% \\ \hline
196 &  77.3 & 19\% & 74\% & 176 & 159.2 & 117.6 & 26.2 & 1.3 &14.0 & 79\% \\ \hline
200 &  83.4 & 18\% & 72\% & 173 & 165.0 & 119.5 & 27.8 & 1.3 &16.4 & 82\% \\ \hline
202 &  40.6 & 17\% & 72\% &  82 &  75.7 &  54.6 & 12.5 & 0.7 & 8.0 & 82\% \\ \hline
206 & 163.9 & 15\% & 70\% & 282 & 274.7 & 193.1 & 47.8 & 2.7 &31.1 & 79\% \\ \hline
207 &  59.4 & 15\% & 70\% & 102 &  99.7 &  70.1 & 17.6 & 1.0 &11.1 & 80\% \\ \hline
\end{tabular}
\end{center}
\caption{\label{tab:pureff}Centre-of-mass energy ($\sqrt{s}$ in GeV),
integrated luminosity ($\cal L$ in pb$^{-1}$), efficiency and purity of the data samples,
number of selected events,
number of expected events from 4-jet WW and background processes
(total and separated by process),
and efficiency of correct pairing of jets to the same W boson.}
\end{table}

The efficiency of the event selection criteria decreases with increasing
centre-of-mass energy. This is primarily due to the `large' angles being reduced
as a result of the increased boost  (becoming lower than the cut value of $100^{\circ}$)
and  `small' angles being increased due to the larger phase-space available
(becoming higher than the cut value of $100^{\circ}$). Much for the same reason, the
efficiency to assign two jets to the same W boson in the selected events increases
slightly with increasing centre-of-mass energy,
in opposition to what would
happen at threshold with the W boson decaying into two back-to-back jets, that would never
be selected to come from the same W boson by the requirement that their interjet angle should be
between $100^{\circ}$ and $140^{\circ}$.

\begin{figure}[ptbh]
\begin{center}
\mbox{\epsfxsize0.8\linewidth\epsffile{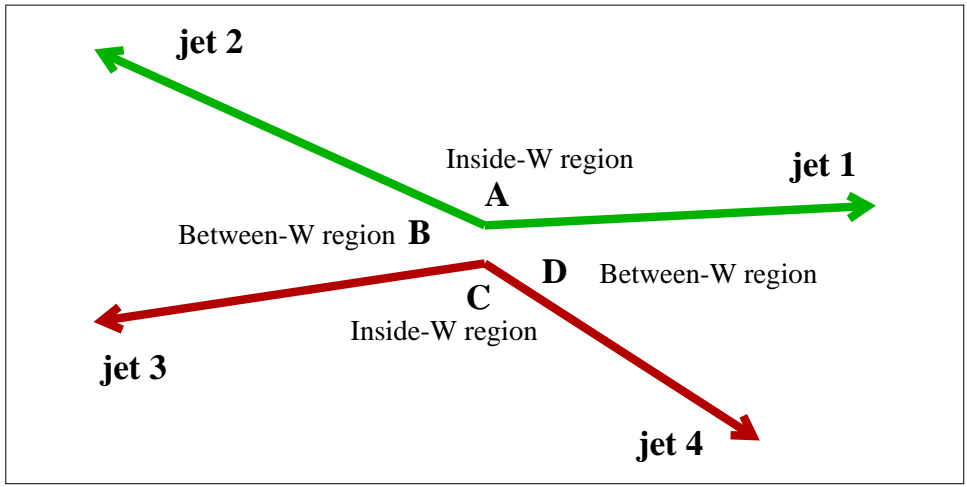}}
\caption[]{Schematic drawing of the angular selection.
}\label{metodo}
\end{center}
\end{figure}

In the following analysis the jets and planar regions are labeled as shown in
Figure~\ref{metodo}:
the planar region corresponding to the smallest jet-jet angle is region B
in the plane made by
jets 2 and 3; the second smallest jet-jet angle corresponds
to the planar region D between jets 1 and 4 in the plane made by these two jets;
the planar region corresponding to the greatest of the large
jet-jet angles in this combination is region A and spans the
angle between jets 1 and 2 in the plane made by these jets; and finally
region C corresponds to the planar region spanned by the second large angle,
between jets 3 and 4 in the plane made by these two jets.
In general, the planar regions are not in the same plane, as the decay planes
of the W bosons do not coincide,
and the
large angles in this combination are not necessarily
the largest jet-jet angles in the event.

\begin{figure}[pbht]
\begin{center}
\mbox{\epsfxsize10cm\epsffile{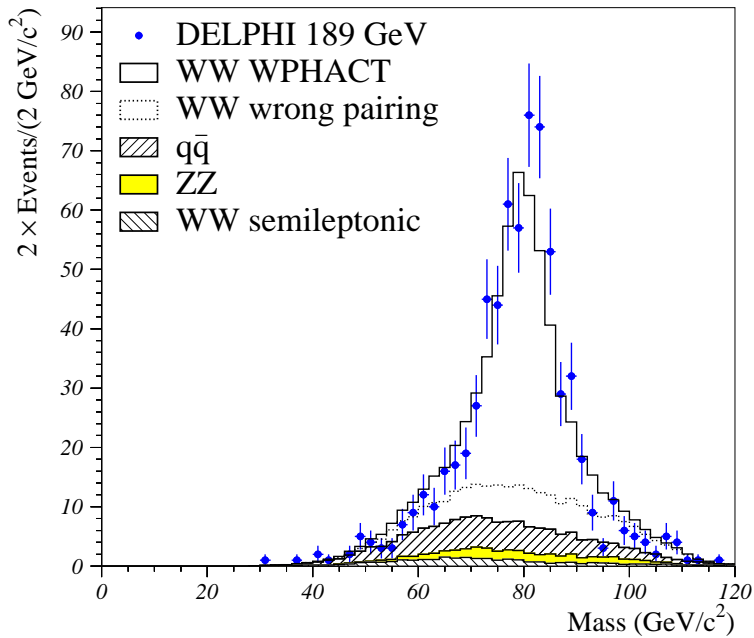}}
\end{center}
\caption[]{Reconstructed dijet masses (after a 4C kinematic fit)
for the selected pairs at 189 GeV
(2 entries per event)(see text).
}\label{wmass}
\end{figure}

The distribution of the reconstructed masses of the jet pairings (1,2) and (3,4),
after applying a 4C kinematic fit requiring energy and momentum conservation,
is shown in Figure~\ref{wmass} (two entries per event).
In the figure, data at 189 GeV (points) are compared
to the expected
distribution from the 4-jet WW signal without CR, plus
background processes, estimated using the simulation (histograms).
The contribution from the 4-jet WW signal simulation is split between the case
in which the two pairs of jets making the large angles actually come from their
parent W bosons and the case in which the jets of a pair come from different W bosons
(mismatch).

\subsection{Particle Flow Distribution}

The particle flow analysis uses 
the number of particles in the Inside-W and the Between-W regions.
An angular ordering of the jets is performed as in Figure~\ref{metodo}.
The two large jet-jet angles in the event are used to define the Inside-W
regions, and the two smallest angles span the Between-W regions, the regions
between the different Ws.

In general, the two W bosons will not decay in the same plane, and this must
be accounted for when comparing the particle production in the Inside-W and
Between-W regions.
So, for each region (A, B, C and D) the particle momenta of all
charged particles are projected
onto the plane spanned by the jets of that region: jets 1 and 2 for region A;
jets 2 and 3 for region B; jets 3 and 4 for region C; jets 4 and 1 for region D.
Then, for each particle the rescaled angle 
$\Phi_{\mathrm{rescaled}}$
is determined as a ratio of two angles:

\begin{equation}
\Phi_{\mathrm{rescaled}} =  \Phi_i/\Phi_r \, , 
\end{equation}
\noindent
when the particle momentum is projected onto the plane of the region $r$.
The angle $\Phi_{i}$ is then the angle between the projected particle momentum
and the first mentioned jet in the definition of the regions given above.
The angle $\Phi_{r}$ is the full opening angle between the jets.
Hence $\Phi_{\mathrm{rescaled}}$  varies between 0 and 1 for the particles
whose momenta are projected between the pair of jets defining the plane.

However, due to the aplanarity of the event about 9\% of the particles in the data
and in the 4-jet WW simulation have projected angles outside all four regions.
These particles were discarded from further analysis.
In the case where a particle could be projected onto more than one region,
with $0<\Phi_{\mathrm{rescaled}}<1$, the solution with the lower momentum
transverse to the region was used.
This happened for about 13\% of the particles in data, after background subtraction,
and
in the 4-jet WW simulation.

%
%

\begin{figure}[p]
\begin{center} \mbox{\epsfxsize10cm\epsffile{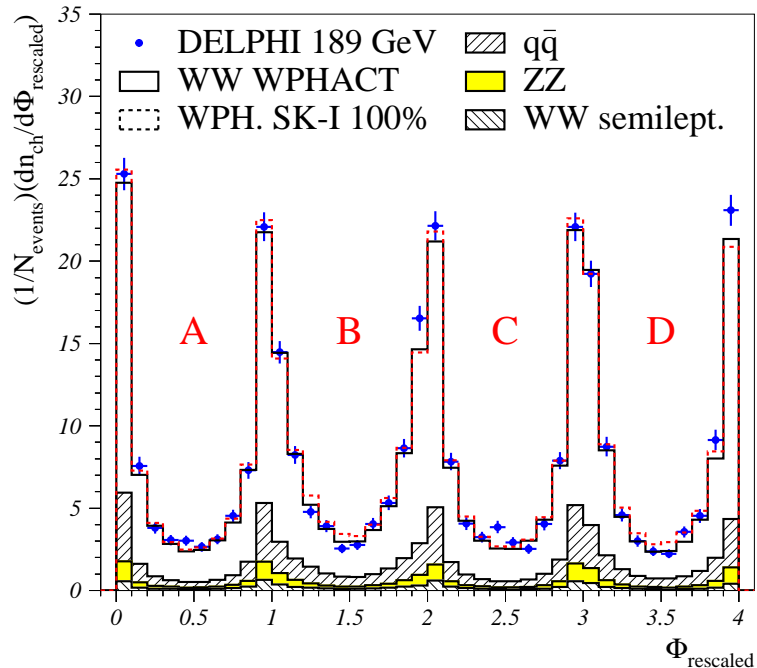}}
\end{center} \caption[]{Normalised charged-particle flow at 189
GeV. The lines correspond to the sum of the simulated 4-jet WW
signal with the background contributions (estimated from DELPHI
MC samples), normalised to the total number of expected events
($\mathrm{N_{events}}$).
The dashed histogram corresponds to the
sum with the simulated 4-jet WW signal generated by WPHACT with 100\%
\SKI.
}\label{rescaled}
\end{figure}

This leads to the normalised particle flow distribution shown in
Figure~\ref{rescaled} at 189 GeV, where the rescaled angle of
region A is plotted from 0 to 1, region B from 1 to 2, region C
from 2 to 3 and region D from 3 to 4. The statistical error on the
bin contents (the average multiplicity per bin of
$\Phi_{\mathrm{rescaled}}$ divided by the bin width) was estimated
using the Jackknife method~\cite{jackknife}, to correctly account
for correlations between different bins.
In this distribution the
regions between the jets coming from the same W bosons
(A and C), and from different W bosons
(B and D), have the same scale and thus can be easily compared.

\begin{figure}[p]
\begin{center}
\mbox{\epsfxsize0.495\linewidth\epsffile{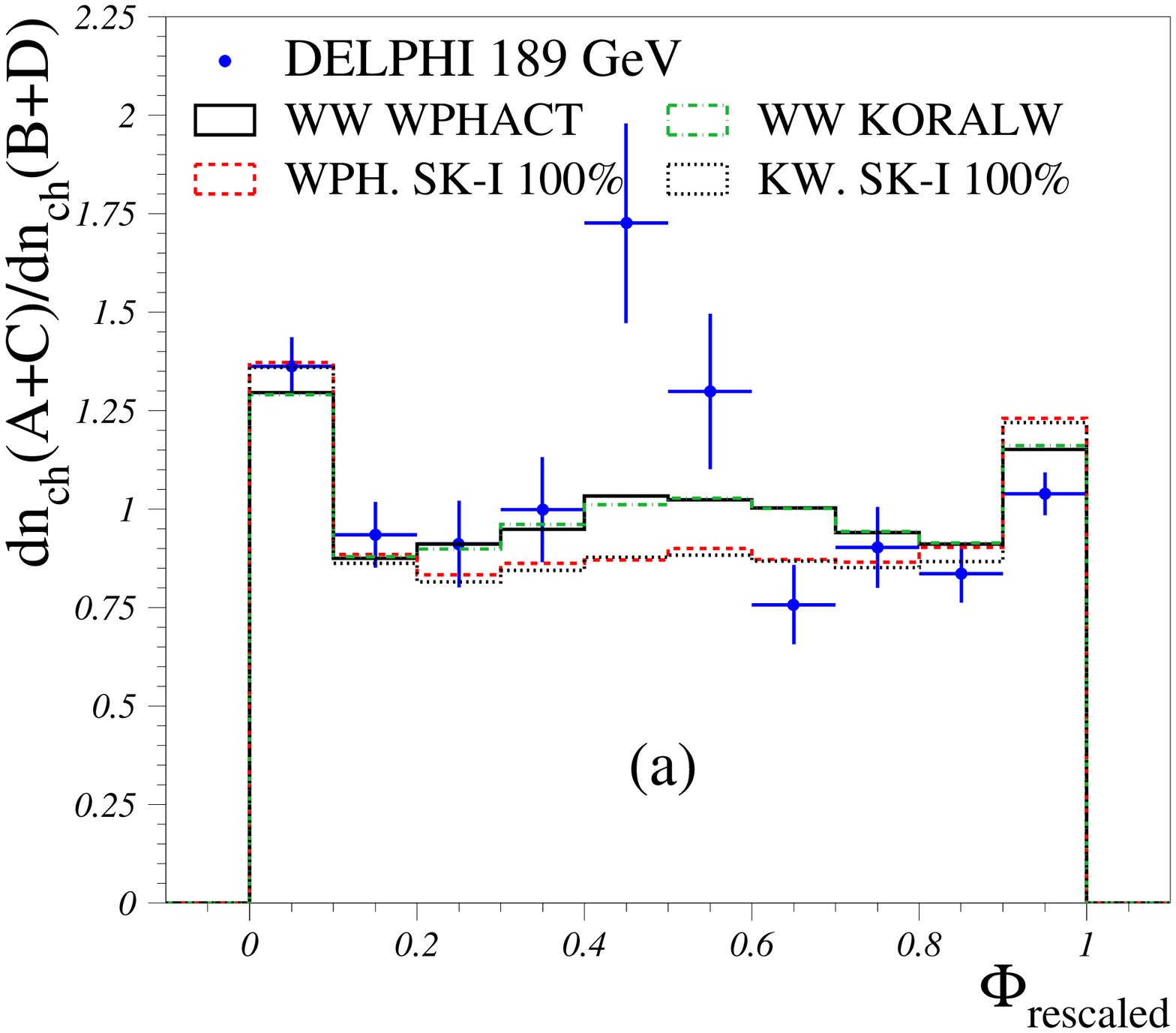}}~\hfill
\mbox{\epsfxsize0.495\linewidth\epsffile{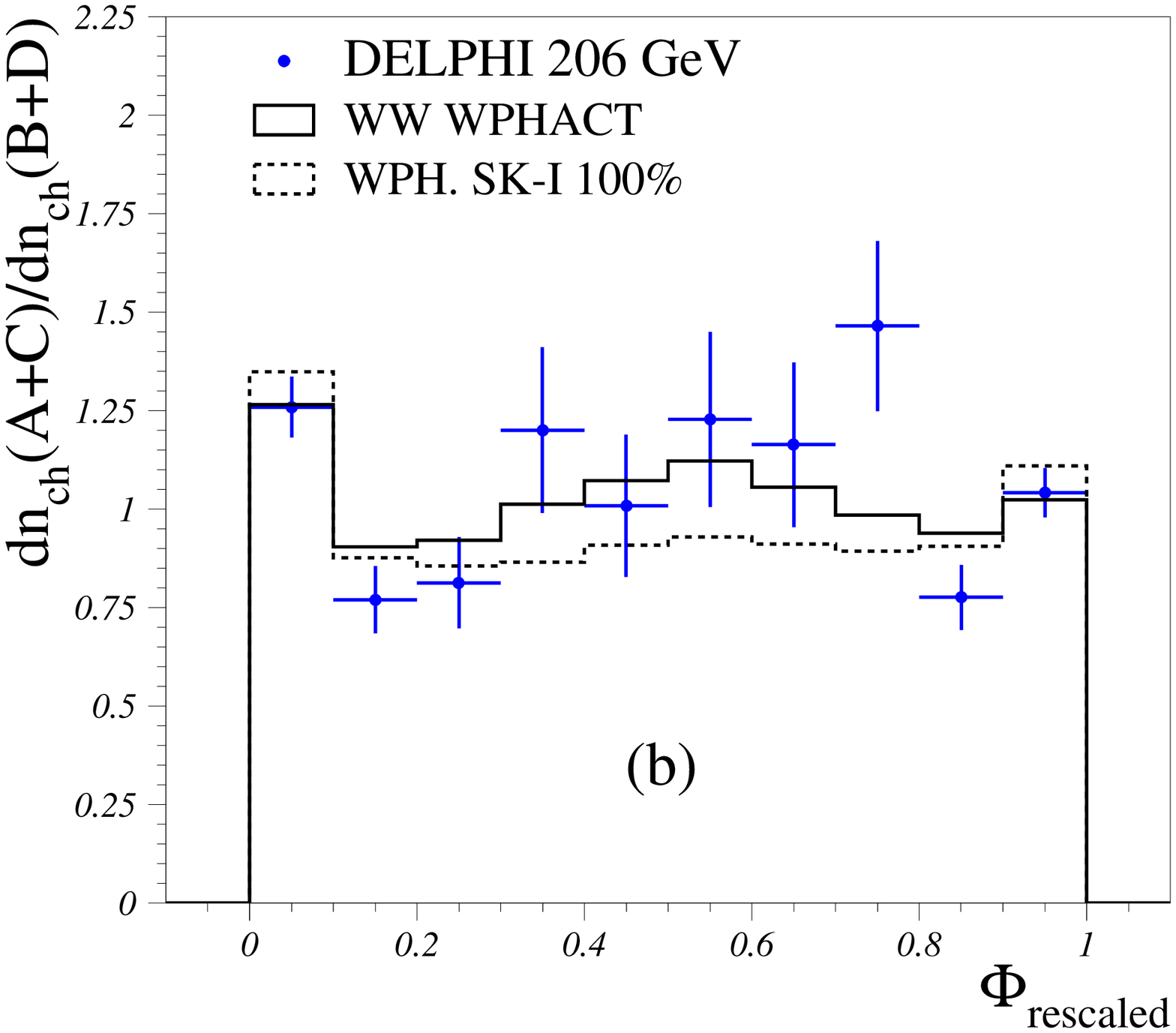}}
\end{center}
\caption[]{The ratio of the particle flow distributions (A+C)/(B+D) at 189 GeV (a)
and at 206 GeV (b).
The data (dots) are compared to WW MC samples generated with
\WPHACT\ (DELPHI samples) and \KORALW\ (Cetraro samples),
both without CR and
implementing the \SKI\ model with 100\% probability of reconnection.
The lines corresponding to \WPHACT\ are hardly distinguishable from the
lines corresponding to \KORALW\ in the same condition of implementation of CR.}
\label{ratio}
\end{figure}

After subtracting bin-by-bin the expected background from the observed
distributions, we define the Inside-W (Between-W) particle flow as
the bin-by-bin sum of regions A and C (B and D).
These distributions are compared by performing the bin-by-bin ratio
of the Inside-W particle flow to the Between-W particle flow.
This ratio of distributions is shown for 189 GeV and 206 GeV
in Figure~\ref{ratio}.
The data points are compared to several fully simulated WW MC
samples with and without CR.

A good agreement was found between the predictions using the \WPHACT\ WW MC
samples and the predictions based on the \KORALW\ WW MC samples, both for the
scenario without CR and for the scenario with CR (\SKI\ model with 100\% probability
of reconnection). For both sets of predictions the regions of greatest difference
between the two scenarios span the rescaled variable $\Phi_{\mathrm{rescaled}}$ from
0.2 to 0.8.

\subsection{Particle Flow Ratio}

After summing the particle flow distributions for regions A and C,
and regions B and D, the resulting distributions are integrated
from 0.2 to 0.8.
The ratio $R$ of the Inside-W to the Between-W particle flow is then defined
as (with $\Phi$ being the rescaled variable $\Phi_{\mathrm{rescaled}}$):

\begin{equation}
 R = \frac{\int_{0.2}^{0.8} dn_{\mathrm{ch}}/d\Phi(A+C) d\Phi
         }{\int_{0.2}^{0.8} dn_{\mathrm{ch}}/d\Phi(B+D) d\Phi} \, . 
\end{equation}

To take into account possible statistical correlations between
particles in the Inside-W
and Between-W regions, the statistical error on this ratio $R$ was
again estimated through the Jackknife method~\cite{jackknife}.

\begin{figure}[hbt]
\begin{center} \mbox{\epsfxsize13cm\epsffile{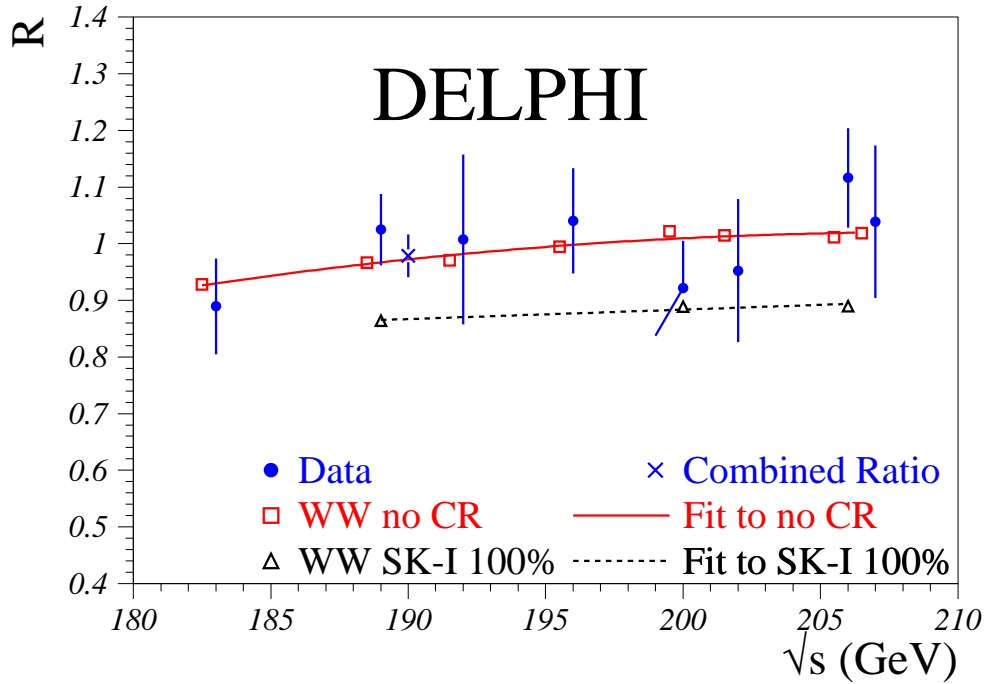}}
\end{center} \caption[]{The ratio $R$ as function of $\sqrt{s}$
for data and MC (DELPHI \WPHACT\ WW samples), and fits to the MC
with and without CR, and the combined ratio after rescaling all
values to $\sqrt{s}=189$~GeV (see text).
The value of the combined ratio
at 189~GeV is shown at a displaced energy (upwards by 1 GeV)
for better visibility, as well as all the values for the MC `WW no CR'
points and
the corresponding fitted curve which are
shown at centre-of-mass energies shifted downwards by 0.5 GeV.
All errors for the MC values
are smaller than the size of the markers.}\label{fit}
\end{figure}

The values for $R$ obtained for the different centre-of-mass energies
are shown in Table~\ref{table:ratio}, and compared to the expectations
from the DELPHI \WPHACT\ WW samples without CR and
implementing the \SKI\ model with 100\% reconnection probability.
These values for data and MC are plotted as function of the
centre-of-mass energy in Figure~\ref{fit}.

\begin{table}
\begin{center}
\begin{tabular}{|c||c|c|c|}
\hline
$\sqrt{s}$ (GeV) & $R_{\mathrm{Data}}$ & $R_{\mathrm{no\: CR}}$
& $R_{\mathrm{\mbox{\SKI}: 100\% }}$ \\ \hline
183 & 0.889 $\pm$ 0.084 & 0.928 $\pm$ 0.005 & -                 \\ \hline
189 & 1.025 $\pm$ 0.063 & 0.966 $\pm$ 0.006 & 0.864 $\pm$ 0.005 \\ \hline
192 & 1.008 $\pm$ 0.150 & 0.970 $\pm$ 0.006 & -                 \\ \hline
196 & 1.041 $\pm$ 0.093 & 0.995 $\pm$ 0.006 & -                 \\ \hline
200 & 0.922 $\pm$ 0.084 & 1.022 $\pm$ 0.007 & 0.889 $\pm$ 0.006 \\ \hline 
202 & 0.952 $\pm$ 0.126 & 1.015 $\pm$ 0.008 & -                 \\ \hline
206 & 1.116 $\pm$ 0.088 & 1.012 $\pm$ 0.008 & 0.889 $\pm$ 0.006 \\ \hline 
207 & 1.039 $\pm$ 0.135 & 1.019 $\pm$ 0.008 & -                 \\ \hline
\end{tabular}
\end{center}
\caption{\label{table:ratio}Values of the ratio $R$ for each energy
(errors are statistical only), and expected values with errors due
to limited statistics of the simulation, all from DELPHI \WPHACT\ WW samples.
}
\end{table}

The changes in the value of $R$ for the MC samples are mainly
due to the different values of the boost of the W systems.
In order to quantify this effect a linear function
$\mathrm{R(\sqrt{s}-197.5)=A+B\cdot (\sqrt{s}-197.5)}$
was fitted to the MC points
with CR (with $\sqrt{s}$ in GeV),
while for the points without CR the quadratic function
$\mathrm{R(\sqrt{s}-197.5)=
\alpha+\beta\cdot (\sqrt{s}-197.5)+\gamma\cdot (\sqrt{s}-197.5)^2}$
was assumed (with $\sqrt{s}$ in GeV),
giving reasonable $\chi^2/d.o.f.$ values. 
The fits yielded the results shown in Table~\ref{table:fit}.

\begin{table}
\begin{center}
\begin{tabular}{|c||c|c|c|c|c|}
\hline
MC Sample & $\chi^2/DF$ & $\alpha$,A       & $\beta$,B                         &
 $\gamma$                       \\ \hline
no CR     & 7.31/5      &$1.001\pm 0.003$ & $(3.20\pm 0.36) \times 10^{-3}$ &
$(-1.35\pm 0.40)\times 10^{-4}$\\ \hline
\SKI\ 100\% & 1.46/1     &$0.880\pm 0.003$ & $(1.68\pm 0.44) \times 10^{-3}$ &
 -                               \\ \hline
\end{tabular}
\end{center}
\caption{\label{table:fit}Results of the fit to the evolution of $R$
with $\mathrm{(\sqrt{s}(\mathrm{GeV})-197.5)}$.}
\end{table}

The MC without CR shows a stronger dependence on $\sqrt{s}$.
The function fitted to this sample was used to rescale the measured values of
$R$ for the data collected at different energies to the energy of 189 GeV,
the centre-of-mass energy at which the combination of the results
of the LEP experiments was proposed in~\cite{review}.
All the rescaled values were combined
with a statistical error-weighted average.
The average of the $R$ ratios rescaled to 189 GeV was found to be

\begin{equation}
\langle R\rangle = 0.979 \pm 0.032 \mathrm{(stat)}. 
\end{equation}

%
%

Performing the same weighted average when using for the rescaling
the fit to the MC with CR, one obtains:

\begin{equation}
\langle R_{\mathrm{CR\: rescale}}\rangle = 0.987 \pm 0.032 \mathrm{(stat)}. 
\end{equation}

Repeating the procedure, but now without rescaling the $R$ ratios,
the result is:

\begin{equation}
\langle R_{\mathrm{no\: rescale}}\rangle = 0.999 \pm 0.033 \mathrm{(stat)}. 
\end{equation}

\subsection{Study of the Systematic Errors in the Particle Flow}

The following effects were studied as sources of systematic uncertainties
in this analysis.



\subsubsection{Fragmentation and Detector response\label{sec441}}

A direct comparison between the particle flow ratios measured in fully
hadronic data and MC samples, $R_{\mathrm{4q\, Data}}$ and
$R_{\mathrm{4q\, MC}}$, respectively, is hampered by the uncertainties
associated with the modelling of the WW fragmentation and the detector
response.
These systematic uncertainties were estimated using mixed
semi-leptonic events. In this technique, two hadronically decaying W bosons
from semi-leptonic events were mixed together to emulate a fully hadronic
WW decay.


\vspace{0.2cm}
\noindent{\bf{Mixing Technique}}
\vspace{0.2cm}

Semi-leptonic WW decays were
selected from the data collected by DELPHI at centre-of-mass
energies between 189 and 206 GeV, by requiring two hadronic jets, a well
isolated identified muon or electron or, in case of a tau candidate, a well
isolated particle, all associated with missing momentum (corresponding to the neutrino)
pointing away from
the beam pipe. A neural network selection,
developed in~\cite{chhiggs}, was used to select the events. The same
procedure was applied to the \WPHACT\ samples fragmented with \PYTHIA\ and
\HERWIG\ at centre-of-mass energies of
189, 200 and 206 GeV and with \ARIADNE\ at 189 and 206 GeV.
The background to this selection was found to be of negligible importance in
this analysis.
 Samples of mixed
semi-leptonic events were built separately at each centre-of-mass energy
for data and Monte
Carlo semi-leptonic samples, following the mixing procedure developed
in~\cite{becww}.

In each semi-leptonic event, the lepton (or tau-decay jet) was stripped off
and the remaining particles constituted the hadronically decaying W boson.
Two hadronically decaying W bosons were then mixed together to emulate a
fully hadronic WW decay. The hadronic parts of W bosons were mixed in such a
way as to have the parent W bosons back-to-back in the emulated
fully hadronic WW decay. To increase the statistics of emulated events,
and profiting from the cylindrical symmetry of the detector along the z axis,
the hadronic parts of W bosons were rotated around the z axis, but were
not moved from barrel to forward regions or vice-versa, as detailed in
the following.

When mixing the hadronic parts of different W events
it was required that the two Ws
had reconstructed polar angles back-to-back or equal within 10 degrees.
 In the latter case, when both Ws are on the same side of the detector,
the z component of the momentum is sign flipped for all the
particles in one of the Ws.

The particles of one W event were then rotated around the beam axis,
in order to have the two Ws also back-to-back in the transverse plane.
Each semi-leptonic event was used in the mixing procedure between 4 and 9 times,
to minimize the statistical error on the particle flow ratio $R$ measured in
the mixed semi-leptonic data sample.



\begin{figure}[hbt]
\begin{center}
\mbox{\epsfxsize10cm\epsffile{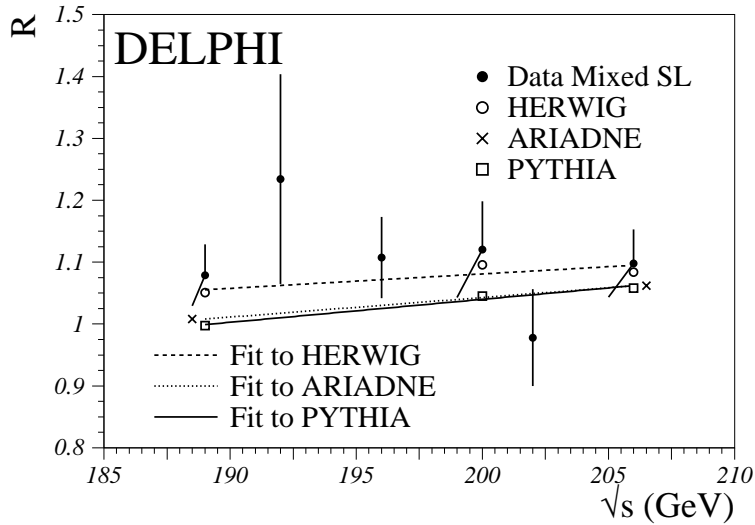}}
\end{center}
\caption[]{The ratio $R_{\mathrm{mixed\, SL}}$ as function of $\sqrt{s}$ for
data and MC, and fits to the MC (see text). The \ARIADNE\ points at
189 GeV and at 206 GeV have
their centre-of-mass energy shifted and
the error bars on data are tilted for readability.
}\label{mix}
\end{figure}

The mixed events were then subjected to the same event selection and particle
flow analysis used for the fully hadronic events.
The particle flow ratios $R_{\mathrm{mixed\, SL\, Data}}$
and $R_{\mathrm{mixed\, SL\, MC}}$ were measured in the mixed
semi-leptonic data and MC samples, respectively, and are plotted as
function of the centre-of-mass energy in Figure~\ref{mix}.

\vspace{0.2cm}

The values of $R_{\mathrm{mixed\, SL}}$ measured in MC show a dependence on
$\sqrt{s}$.
This effect is quantified by performing linear fits to the points measured
with \PYTHIA, \ARIADNE\ and \HERWIG, respectively.
The differences between the measured
slopes were found to be small. The function fitted to the \PYTHIA\ points
was used to rescale the values of $R$ measured in data at different energies
to 189~GeV. The rescaled values were then combined using as
weights the scaled statistical errors.
The weighted average $R$ at 189 GeV for the mixed semi-leptonic
events built from data was found to be

\begin{equation}
\langle R_{\mathrm{mixed \, SL \, Data}} \rangle = 1.052 \pm 0.027 {\mathrm{(stat)}}. 
\end{equation}

For each MC sample, the ratio
$ R_{\mathrm{mixed \, SL \, Data}}
/ R_{\mathrm{mixed \, SL \, MC}}$
was used to calibrate the particle flow
ratio measured in the corresponding fully hadronic sample,
$R_{\mathrm{4q\, MC}}$, to compare it to the ratio measured in the data,
$\langle R_{\mathrm{4q\, Data}} \rangle$.
%
%
%
%
The correction factor
$ R_{\mathrm{mixed \, SL \, Data}}
/ R_{\mathrm{mixed \, SL \, MC}} $
was computed from the values of $R$ rescaled to 189 GeV,
calculated from the fits to the mixed
semi-leptonic samples built from the data and the MC.
The values for $R_{\mathrm{mixed \, SL \, MC}}$ are presented in
Table~\ref{table:mcmix}, for the different models,
along with the calibrated values of $R_{\mathrm{4q}}$ for the same models.

\begin{table}
\begin{center}
\begin{tabular}{|c|c|c|c|}
\hline
MC sample                    &  \PYTHIA  &  \ARIADNE &  \HERWIG \\ \hline
$R_{\mathrm{\, mixed \, SL\, Data}} / R_{\mathrm{\, mixed \, SL\, MC}}$
                             &  $1.053$  &  $1.044$  &  $0.997$ \\
$R_{\mathrm{4q\, MC}}^{\mathrm{Calibrated}}$
                             &  $1.018$  &  $1.011$  &  $1.004$ \\ \hline
\end{tabular}
\end{center}
\caption{\label{table:mcmix}Ratio of data to MC fitted values of $R$ in mixed
semi-leptonic samples, used to calibrate the $R_{\mathrm{4q\, MC}}$ values for
different models (upper line), and calibrated values of
$R_{\mathrm{4q\, MC}}$.
All values were computed at $\sqrt{s}=189$~GeV.}
\end{table}

The calibration factors differ from unity by less than 6\%, and the
largest difference of the calibrated $R_{\mathrm{4q\, MC}}$ values
when changing the
fragmentation model, $0.014$, was considered as
an estimate of the systematic error due to
simulation of the fragmentation and of the detector response,
and was added in quadrature to the systematic error.
The error in the calibrated $R_{\mathrm{4q\, MC}}$ values due to the
statistical error on $\langle R_{\mathrm{mixed \, SL \, Data}} \rangle$
value used for the calibration, $0.026$,
was also added in quadrature to the systematic error.

\subsubsection{Bose-Einstein Correlations\label{sec442}}

Bose-Einstein correlations (BEC) between identical pions and kaons
are known
to exist and were established and studied in Z hadronic decays
in~\cite{becZ}. They are expected to exist with a similar
behaviour in the W hadronic decays, and this is studied
in~\cite{becww}. They are implemented in the MC simulation
samples with BEC via the BE$_{32}$ model of LUBOEI~\cite{bec3}, which was
tuned to describe the DELPHI data in~\cite{becww}. However, the
situation for the WW (ZZ) fully hadronic decays is not so clear,
i.e. whether there are correlations only between pions and
kaons coming from
the same W(Z) boson or also between pions and kaons
from different W(Z)
bosons. The analyses of Bose-Einstein correlations between
identical particles coming from the decay of different W bosons do
not show a significant effect~\cite{bec} for three of the LEP
experiments, whereas for DELPHI, an effect was found at the level
of 2.4 standard deviations~\cite{becww}.
Thus, a comparison
was made between
the \WPHACT\ samples without CR and with BEC only between the identical
pions coming from the same W boson (BEC only inside), to the samples 
without CR and with BEC allowed for all the particles stemming
from both W bosons, implemented with the BE$_{32}$ variant of the LUBOEI model
(BEC all).
The $R$
values were obtained at each centre-of-mass energy, after which a
linear fit was performed for each model to obtain a best
prediction at 189 GeV. The fit values were found to be in
agreement to the estimate at 189 GeV alone, and for simplicity
this estimate was used.
The measurement of BEC from DELPHI of 2.4 standard deviations above zero
(corresponding to BEC only inside), was used to interpolate the range of 4.1 standard
deviations of separation between BEC only inside and BEC all. To include the error on the
measured BEC effect, one standard deviation was added to the effect before the
interpolation.
The difference in the estimated values of $R$ at
$\sqrt{s}=189$~GeV, 
between the model with BEC only inside and the model with partial BEC all
(at the interpolated point of 3.4/4.1),
-0.013, was
added in quadrature to the systematic error.


\subsubsection{$\mathrm{q\bar q}(\gamma)$ Background Shape}

The fragmentation effects, in the shape of the $\mathrm{q\bar q}(\gamma)$
background, were
estimated by comparing the values of $R$ obtained when the subtracted
$\mathrm{q\bar q}(\gamma)$ sample was fragmented with \ARIADNE\ instead of
\PYTHIA\ at the centre-of-mass energy of 189~GeV, and the difference, 0.003,
was added in quadrature to the systematic error.

%

\subsubsection{$\mathrm{q\bar q}(\gamma)$ and ZZ Background Contribution}

At the centre-of-mass energy of 189~GeV,
the $\mathrm{q\bar q}(\gamma)$ cross-section in the 4-jet region
is poorly known, due to the difficulty in isolating the
$\mathrm{q\bar q}(\gamma) \rightarrow$~4-jet signal
from other 4-jet processes such as WW and ZZ. The study
performed in~\cite{wxsec} has shown that the maximal difference in
the estimated $\mathrm{q\bar q}(\gamma)$ background rate is 10\%
coming from changing from
\PYTHIA\ to \HERWIG\ as the hadronization model, with the
\ARIADNE\ model giving intermediate results.
Conservatively, at each
centre-of-mass energy a variation of 10\%
on the $\mathrm{q\bar q}(\gamma)$ cross-section was assumed,
and the largest shift in $R$, 0.011, was added in quadrature to
the systematic error.


The other background process considered is the Z pair production.
The Standard Model predicted cross-sections are in agreement with the
data at an error level of 10\%~\cite{zzxsec}.
The cross-section was thus varied by $\pm 10\%$
at each energy and the effect in
$R$ was found to be negligible.
%

\subsubsection{Evolution of $R$ with Energy}
The $R$ ratios were rescaled to $\sqrt{s}=189$~GeV using the fit to the
MC without CR, however the correct behaviour might be given by the MC with CR.
Hence, the difference of 0.009 between the R values
obtained using the two rescaling methods, using MC without CR
$\langle R\rangle$ and with CR $\langle R_{\mathrm{CR\: rescale}}\rangle$,
was added in quadrature to the systematic error.

\subsection{Results of the Particle Flow Analysis}

The final result for the average of the ratios $R$ rescaled
to 189 GeV is

\begin{equation}
\langle R\rangle = 0.979 \pm 0.032 \mathrm{(stat)} \pm 0.035 \mathrm{(syst)}.
\end{equation}

In order to facilitate comparisons between the four LEP
experiments, this value can be normalised by the one determined
from simulation samples produced with the full detector simulation
and analysed with the same method. The LEP experiments agreed to
use for this purpose the Cetraro \PYTHIA\ samples. These events
were generated with the ALEPH fragmentation tuning but have been
reconstructed with the DELPHI detector simulation and analysed
with this analysis.
The values of the $R$ ratios obtained from the
Cetraro samples at 189 GeV, calibrated using the mixed semi-leptonic
events from these samples,
 are given in Table~\ref{table:cetro}.

\begin{table}
\begin{center}
\begin{tabular}{|c||c|c|c|}
\hline
MC Sample            & $R$               \\ \hline
\PYTHIA\  no CR      & $1.037 \pm 0.004$ \\ \hline
\PYTHIA\  \SKI\ 100\% & $0.917 \pm 0.003$ \\ \hline
\ARIADNE\ no CR      & $1.053 \pm 0.004$ \\ \hline
\ARIADNE\ AR2        & $1.021 \pm 0.004$ \\ \hline
\HERWIG\  no CR      & $1.059 \pm 0.004$ \\ \hline
\HERWIG\  1/9 CR     & $1.040 \pm 0.003$ \\ \hline
\end{tabular}
\end{center}
\caption{\label{table:cetro}$R$ ratios for the Cetraro samples at 189 GeV, calibrated with
the mixed semi-leptonic events.}
\end{table}

The value of $\langle R\rangle$ measured from data is  between the
expected $R$ ratios from \PYTHIA\ without CR and with the
\SKI\ model with 100\% fraction of reconnection.
The error of this measurement is larger than the difference
between the values of $R$ from \ARIADNE\ samples without and with CR, and than the
difference between values of $R$ from
the \HERWIG\ samples without CR and with 1/9 of reconnected events.

The following normalised ratios are obtained for the sample
without CR and implementing the \SKI\ model with 100\% CR
probability, respectively:

\begin{eqnarray}
\label{eq_1}
r^{\mathrm{data}}_{\mathrm{no\: CR}} = \frac{\langle R\rangle_{\mathrm{data}}}{R_{\mathrm{no\: CR}}}
= 0.944 \pm 0.031 \mathrm{(stat)} \pm 0.034 \mathrm{(syst)}, \\
r^{\mathrm{data}}_{\mathrm{CR}}      =
\frac{\langle R\rangle_{\mathrm{data}}}{R_{\mathrm{CR}}} =
1.067 \pm 0.035 {\mathrm{(stat)}} \pm 0.039 {\mathrm{(syst)}}. \end{eqnarray}

\noindent
In the above expressions,
the statistical errors in the MC predicted values were propagated
and added quadratically
to the systematic errors on the ratios.

\begin{figure}[hbt]
\begin{center} \mbox{\epsfxsize13cm\epsffile{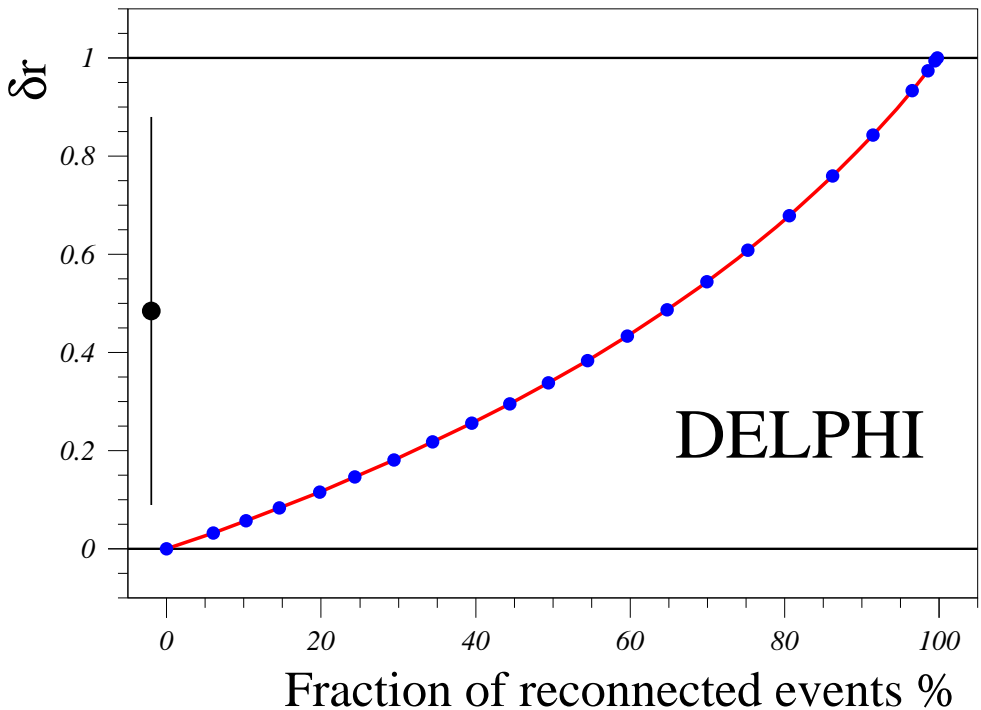}}
\end{center} \caption[]{Comparison of the
measurement of the $\delta r$ observable to the predictions from
the \SKI\ model as a function of the fraction of reconnected events.
\label{sk1pflow-dr}} \end{figure}

It is also possible to define the following quantity,
taking the predictions for $R_{\mathrm{CR}}$ and $R_{\mathrm{no\: CR}}$
at $\sqrt{s}=189$~GeV
from the \PYTHIA\ samples in Table~\ref{table:cetro},

\begin{equation}
\label{eq_3}
\delta r=\frac{\langle R_{\mathrm{data}}\rangle -R_{\mathrm{no\: CR}}
             }{R_{\mathrm{CR}}-R_{\mathrm{no\: CR}}                  }
= 0.49 \pm 0.27 {\mathrm{(stat)}} \pm 0.29 {\mathrm{(syst)}}\, ,
\end{equation}

\noindent
from which it can be concluded that the
measured $\langle R_{\mathrm{data}}\rangle$ is compatible with 
intermediate
probability of CR, and differs from the CR in the \SKI\ model at 100\%
at the level of 1.3 
standard deviations.
The ability to distinguish between these two models can be computed from the inverse of the sum
in quadrature of the statistical and
systematic errors; it amounts to be 2.5 
standard deviations.
In Figure~\ref{sk1pflow-dr} the result of $\delta r$ is compared to the
predicted values, in the scope of the \SKI\ model, as a funtion of the
fraction of reconnected events.

\begin{figure}
\begin{center}
\mbox{\epsfxsize10cm\epsffile{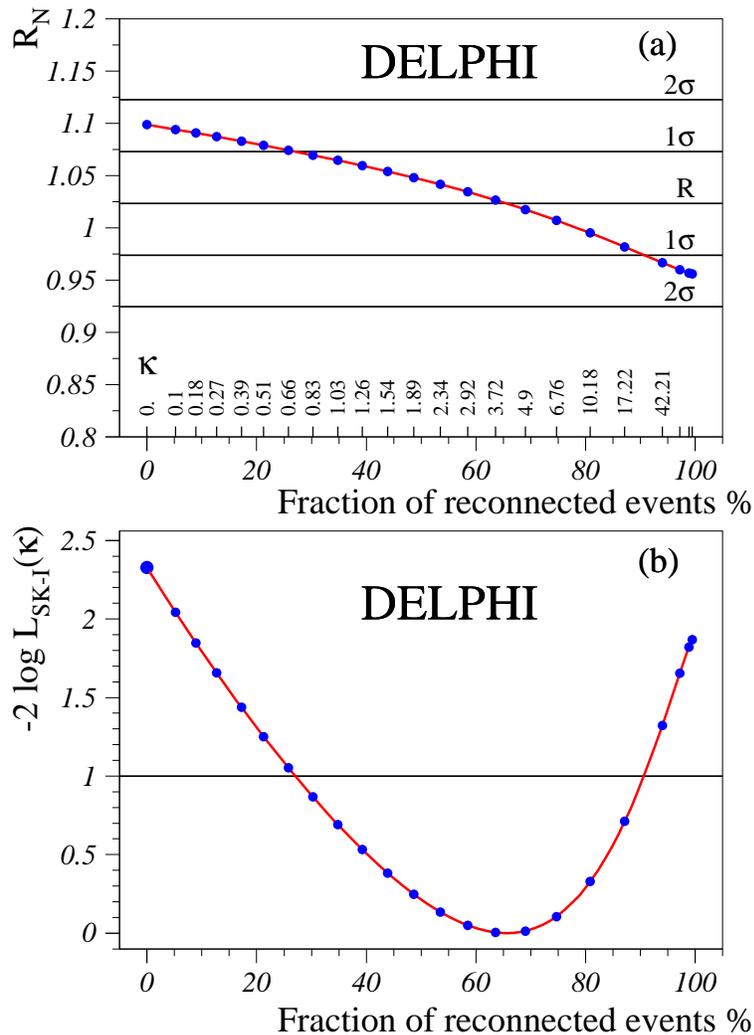}}
\end{center}
\caption[]{a) Estimated ratio $R_N$ at 200~GeV
plotted as a function of different $\kappa$ values (top scale), or as
function of
the corresponding reconnection probabilities (bottom scale), compared
to $\langle R\rangle$ measured from data after rescaling to 200~GeV
(horizontal lines marked with R for the value and with $1\sigma$($2\sigma$) for
the $\langle R\rangle$ value added/subtracted by one(two) standard deviations);
the last three marks on the x axis, close to 100\% of
reconnection probability, correspond respectively to the values
$\kappa=100,300,800$; b) corresponding log-likelihood curve for the
comparison of the estimated values ($R_N$) with the data ($\langle R\rangle$).
\label{sk1pflow}} \end{figure}

The result for the value of $\langle R\rangle$ can also be used to test for consistency
with the \SKI\ model as a function of
$\kappa$ and a log-likelihood curve was obtained.
This also facilitates combination with the result obtained in the analysis in the
following section,
and  for this reason the value of $\langle R\rangle$ is
rescaled with \PYTHIA\ without CR to a centre-of-mass energy of 200 GeV:
the value obtained at 200 GeV is $\langle R\rangle(200\;{\mathrm{GeV}})=1.024\pm0.050$.
The values obtained for the predicted ratios $R_N$ at 200 GeV and the
log-likelihood curve, as a function of $\kappa$, are shown in
Figure~\ref{sk1pflow}.
The value of $\kappa$ most compatible with the data
within one standard deviation is

\begin{equation}
\label{eq_12}
\kappa_{\mathrm{\mbox{\SKI}}} = 4.13^{+20.97}_{-3.46}\, .
\end{equation}

\section{Different \boldmath{$\mw$} Estimators as Observables}
\label{section:analysismw}

It has been shown~\cite{JDH_MWnote} that the $\mw$ measurement inferred from
hadronically decaying
$\mathrm{W^{+}W^{-}}$ events at LEP-2, by the method of direct reconstruction,
is influenced by CR effects, most visible when changing the value
of $\mathrm{\kappa}$ in the \SKI\ model.
For the $\mathrm{M_W(4q)}$ estimator
within DELPHI this is shown in~\cite{paper189}.
Other published $\mw$ estimators in LEP experiments
are equally sensitive to $\mathrm{\kappa}$~\cite{LEPEWmass}.

To probe this sensitivity to CR effects, alternative estimators
for the $\mw$ measurement were designed which
have different sensitivity to $\mathrm{\kappa}$. In the following,
the standard estimator and two alternative estimators, studied
in this paper, are presented. The standard estimator corresponds
to that previously used in the measurement of the W mass by DELPHI~\cite{paper189}.
Note that in the final DELPHI W mass analysis~\cite{wmass} results are given
for the standard and hybrid cone estimators, with the hybrid cone estimator used
to provide the primary result.
The data samples, efficiencies and purities for the
analysis corresponding to the standard estimator are provided in~\cite{paper189,wmass}.

\begin{itemize}

\item {\bf The standard \boldmath{$\mw$} estimator} :

This estimator is described in ~\cite{paper189} and was optimised
to obtain the smallest statistical uncertainty for the W mass
measurement. It results in an event-by-event likelihood
${\mathrm L}_i (\mw)$ for the parameter $\mw$.

\item {\bf The momentum cut \boldmath{$\mw$} estimator} :

For this alternative $\mw$ estimator the event selection
was performed in exactly the same way as for the standard $\mw$
estimator.
The particle-jet association was also taken from this analysis.
However, when reconstructing the event for the $\mw$ extraction a
tighter track selection was
applied. The momentum and energy of the jets were calculated only from
those tracks having a momentum higher than a certain $\Pcut$ value.
An event-by-event likelihood $\mathrm{L}^{\Pcut}_i(\mw)$ was
then calculated.

\item {\bf The hybrid cone \boldmath{$\mw$} estimator} :

In this second alternative $\mw$ estimator the reconstruction of the
event
is the same as for the standard analysis, except when calculating the jet
momenta used for the $\mw$ extraction.

\bfg{tbhp}
  \includegraphics[height=8.5cm]{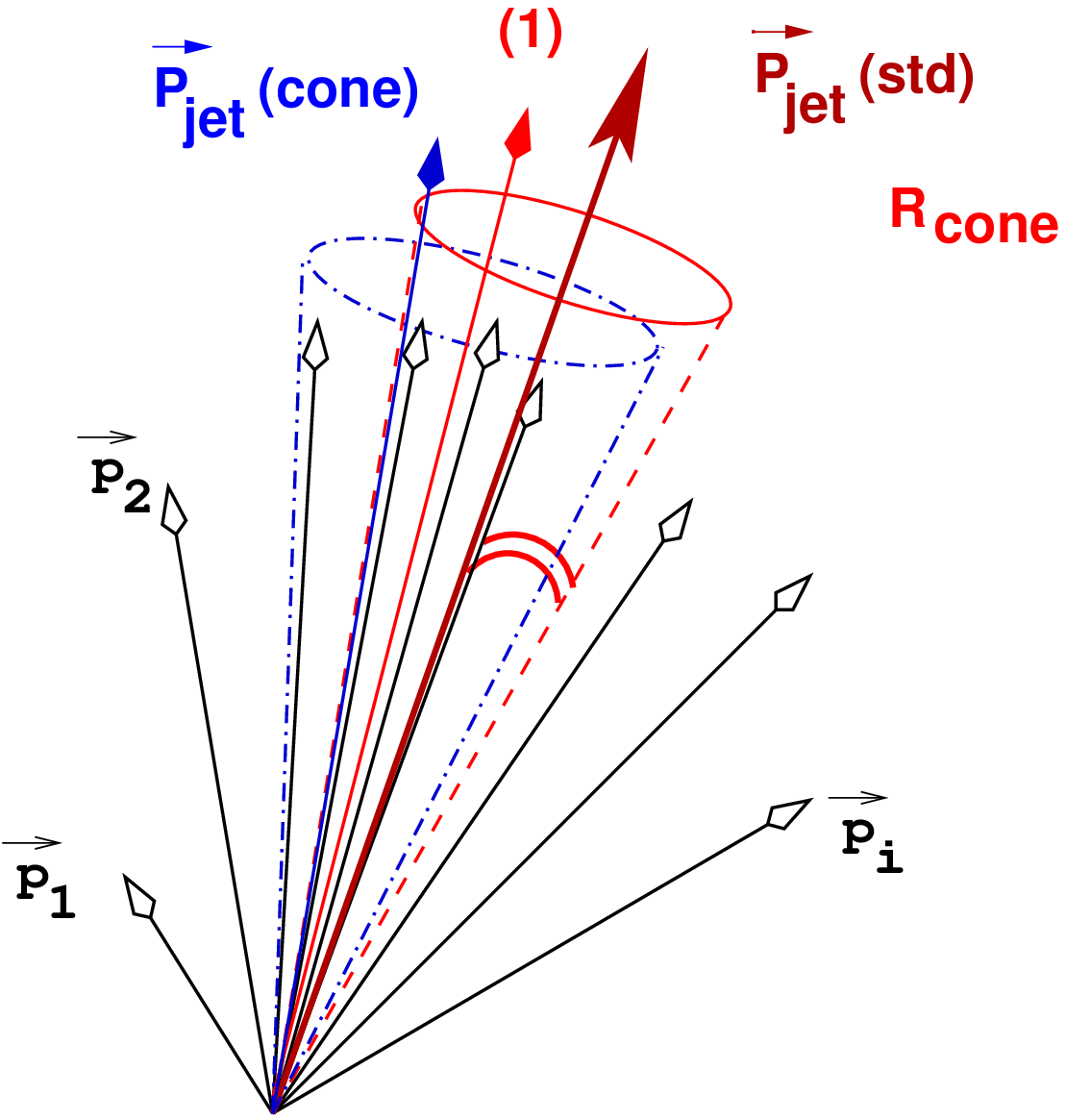}
  \caption{Illustration of the iterative cone algorithm within a predefined
jet as explained in the text.}
\efg{conemw}

An iterative procedure was used within each jet (defined by the
clustering algorithm used in the standard analysis) to find a stable
direction of a cone excluding some
particles in the calculation of the jet momentum, illustrated in
Figure~\rf{conemw}.
Starting with the direction of the original jet
${\vec{p}}^{\mathrm{\, jet}}_{\std}$
, the jet direction
was recalculated (direction (1) on the Figure) only from those
particles which have an opening angle
smaller than $\Rcone$ with this original jet. This process was
iterated
by constructing a second cone (of the same opening angle) around this new
jet direction
and the jet direction was recalculated again.
The iteration was continued until a stable jet
direction 
${\vec{p}}^{\mathrm{\, jet}}_{\cone}$ was found.
The jet momenta obtained,
${\vec{\mathrm p}}^{\mathrm{\, jet}}_{\mathrm{cone}}$, were rescaled to
compensate
for the lost energy of particles outside the stable cone,
\begin{equation}
\vec{\mathrm p}^{\, \mathrm{jet}}_{{\cone}} \rightarrow
\vec{\mathrm p}^{\, \mathrm{jet}}_{\cone} \cdot
\frac{\mathrm{E^{jet}}}{\mathrm{E^{jet}}_{\cone}} \ . 
\end{equation}
The energies of the jets were taken to be the same as those
obtained with the standard clustering algorithm
($\mathrm{E^{\, jet}}_{\cone} \rightarrow
  \mathrm{E^{\, jet}}$).
This was done to increase the correlation of this estimator with the
standard one. The rescaling was not done
for the $\Pcut$ estimator as it will be used in a
cross-check observable with different systematic properties.
Again the result is an event-by-event likelihood
$\mathrm{{L}^{\Rcone}_i(\mw)}$.

\end{itemize}

Each of these previously defined $\mw$ likelihoods had to be
calibrated.
The slope of the linear calibration curve for the $\mw$
estimators is tuned to be unity, therefore
only a bias correction induced by the
reconstruction method has to be applied. This bias is estimated with the
nominal \WPHACT\ Monte
Carlo events and the dependence on the value of $\mathrm{\kappa}$ is
estimated with the
\EXCALIBUR\ simulation.
It was verified for smaller subsets that the results using these
large \EXCALIBUR\ samples and the samples generated with \WPHACT\ are compatible.
Neglecting the possible
existence of Colour Reconnection (CR) in the Monte Carlo simulation results in
event likelihoods
${\mathrm L}_i(\mw |\mbox{event without CR})$, while
${\mathrm L}_i(\mw |\mbox{event with CR})$ are the
event likelihoods obtained
when assuming the hypothesis that events do reconnect (100\% CR in the scope
of the \SKI\ model).
To construct the event likelihoods for intermediate CR
(values of $\mathrm{\kappa}$ larger than 0) the
following weighting formula is used :
\begin{equation}
{\mathrm L}_i(\mw |\kappa) = [ 1-{\cal P}_i(\kappa) ] \cdot
{\mathrm L}_i(\mw |\mbox{event without CR}) + {\cal P}_i(\kappa) \cdot
{\mathrm L}_i(\mw |\mbox{event with CR})
\end{equation}
where $\mathrm{{\cal P}_i(\kappa)}$ is defined in Equation~\ref{eq:sk1}.
The combined likelihood is produced for the event sample;
the calibrated values for $\mw(\kappa)$ were
obtained for different values of $\kappa$ using the maximum likelihood
principle.
In Figure~\rf{sk1curves} the difference
$\mathrm{d\mw(\kappa) = \mw(\kappa)-\mw(\kappa=0)}$ or the influence
of $\mathrm{\kappa}$ on
the bias of the $\mw$ estimator is presented as function of
$\mathrm{\kappa}$.

\bfg{htbp}
  \includegraphics[width=\linewidth]{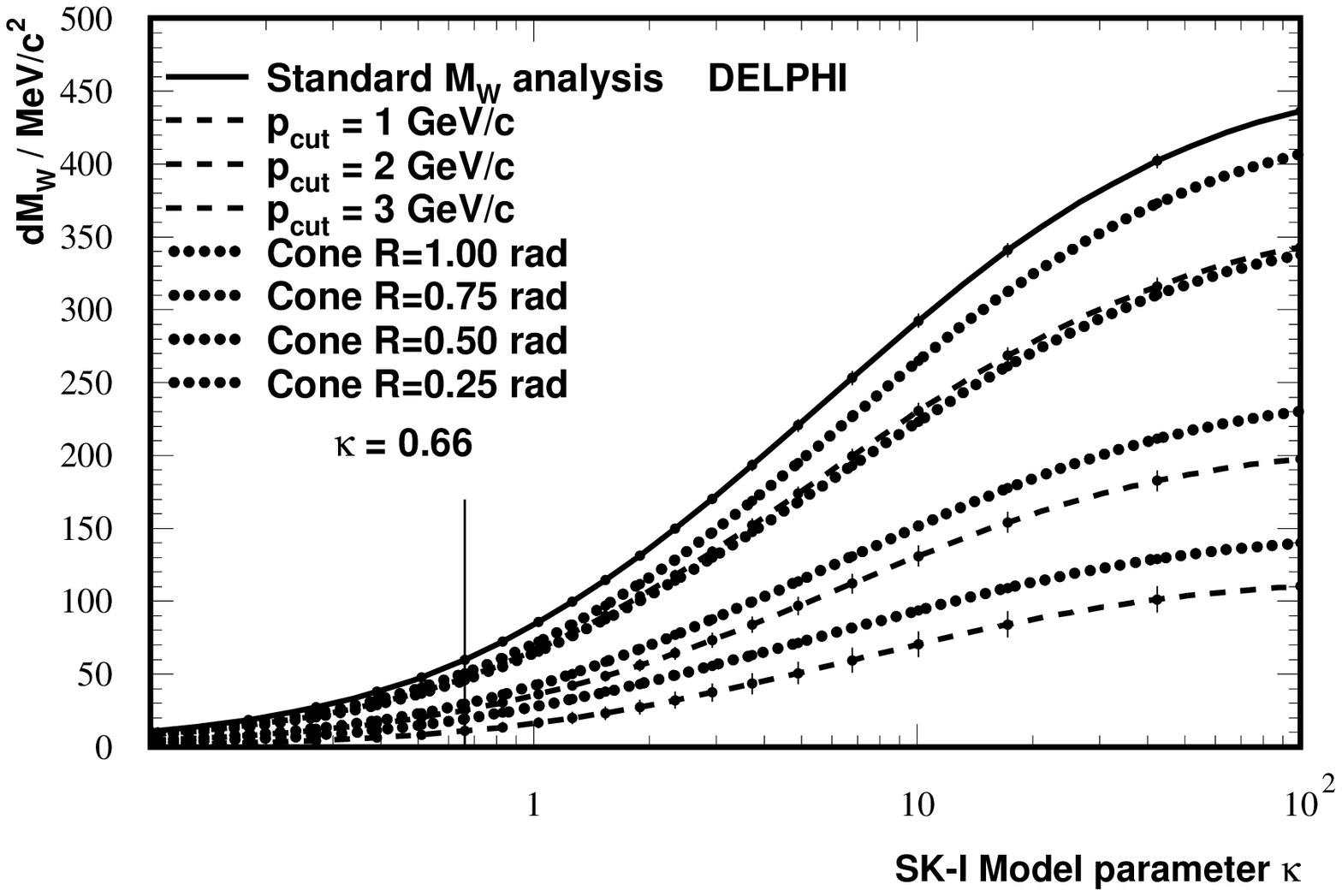}
  \caption{The difference d$\mw(\kappa)=\mw(\kappa)-\mw(\kappa=0)$ is presented as a function
of $\kappa$, for different $\mw$ estimators. The curve for the standard
$\mw$ estimator is the curve at the top
.
The curves obtained with the hybrid cone analysis for different
values of the cone opening angle, starting from the top with 1.00 rad down to
0.75 rad, 0.50 rad and 0.25 rad are indicated with dotted lines
.
The curves obtained with the momentum cut analysis for different values of
$\Pcut$, starting from the top with 1 GeV/c, down to 2 GeV/c and 3 GeV/c
are dashed
. The vertical line indicates the
value of $\kappa$ preferred by the \SKI\ authors~\cite{sk1} and commonly used
to estimate systematic uncertainties on measurements using
$\epem \rightarrow \mathrm{W^{+}W^{-} \rightarrow q_1\bar{q}_2q_3\bar{q}_4} $ events. }
\efg{sk1curves}

The uncertainty on this difference is estimated with the Jackknife
method~\cite{jackknife} to take the correlation
between $\mathrm{\mw(\kappa)}$ and $\mathrm{\mw(\kappa=0)}$ into account.
It was observed from simulations
that the estimators dependency on $\kappa$, for $\kappa$ below about 5,
was not significantly different in the centre-of-mass range between 189
and 207 GeV. Therefore in the determination of $\kappa$ the dependency at
200 GeV was taken as default for all centre-of-mass energies.
This value of centre-of-mass energy is close to the integrated luminosity
weighted centre-of-mass energy of the complete data sample, which is 197.1 GeV.

When neglecting the information content of low momentum
particles or when using the hybrid cone algorithm, the influence of Colour
Reconnection on the $\mw$ estimator is decreased. The dependence
$\mathrm{\frac{\partial \mw}{\partial \kappa}}$
of the estimator to $\mathrm{\kappa}$ is decreased when increasing the value
of $\Pcut$
or when working with smaller cone opening angles $\Rcone$.

\subsection{The Measurement of \boldmath{$\mathrm{\kappa}$}}
\label{section:meask}

The observed difference $\Delta \mw(\std,i) = \mw^{\std}-\mw^i$
in the event sample, where $i$
is a certain
alternative analysis, provides a measurement
of $\mathrm{\kappa}$. When both estimators $\mw^{\std}$ and
$\mw^{i}$ are calibrated in the same hypothesis of
$\mathrm{\kappa}$, 
the expectation values of $\Delta \mw(\std,i)$
will be invariant under a change of $\Pcut$ or $\Rcone$.
%

When neglecting part of the information content of the events in these
alternative $\mw$ analyses, by increasing $\Pcut$ or
decreasing $\Rcone$, the statistical uncertainty on the
value of the $\mw$ estimator is increased.
Therefore a balance must be found between the
statistical precision on $\Delta \mw(\std,i)$
and the dependence of this
difference to $\mathrm{\kappa}$ in order to obtain the largest sensitivity for
a $\mathrm{\kappa}$ measurement. This optimum was found using the Monte Carlo
simulated events and assuming that the data follow the $\mathrm{\kappa=0}$
hypothesis, resulting in the
smallest expected uncertainty on the estimation of $\kappa$.

%
%


For the $\Pcut$ analysis an optimal sensitivity was found when
using the difference $\dmwp$ with
$\Pcut$ equal to 2~GeV/c or 3~GeV/c.
Even more information about $\mathrm{\kappa}$ could be extracted from the
data, when using the difference $\dmwr$, which was
found to have an optimal sensitivity around $\Rcone=0.5$ rad.
No significant improvement in the sensitivity was found when combining the
information from these two observables. Therefore the best measure of
$\mathrm{\kappa}$ using this method is extracted from the
$\dmwrcn$ observable. Nevertheless,
the $\dmwpct$ observable was studied as a
cross-check.

\subsection{Study of the Systematic Errors in the \boldmath{$\Delta\mw$} Method}
\label{section:syst}

The estimation of systematic uncertainties on the observables
$\Delta \mw(\std,i)$
follows similar methods to those used within the $\mw$ analysis.
Here the double difference is a measure of the systematic uncertainty between
Monte Carlo simulation (`MC') and real data (`DA'):

\begin{equation}
\Delta_{\mathrm{syst}}(\mathrm{MC,DA}) =
\left| [ \mw^{\std}(\mathrm{MC}) - \mw^{\std}(\mathrm{DA}) ] -
       [ \mw^{i}(\mathrm{MC}) - \mw^{i}(\mathrm{DA}) ] \right|
\label{doublediff}
\end{equation}

\noindent
where $i$
 is one of the alternative $\mw$ estimators.
The systematic error components are described below and summarised in
Table~\ref{table:syst2}.

\subsubsection{Jet Reconstruction systematics with MLBZs}

A novel technique was proposed in~\cite{mlbz} to study systematic
uncertainties on jet reconstruction and fragmentation in W physics
measurements with high statistical precision through the use of
Mixed Lorentz Boosted $\Z$ events (MLBZs). The technique is similar to the
one described in section~\ref{sec441}.
The main advantage of this method was that Monte Carlo simulated jet
properties in $\mathrm{W^+W^-}$ events could be directly compared with
the corresponding ones from real data using the large 
$\Z$ statistics.

\bfg{pht}
\vspace{0.5cm}
  \includegraphics[height=14cm]{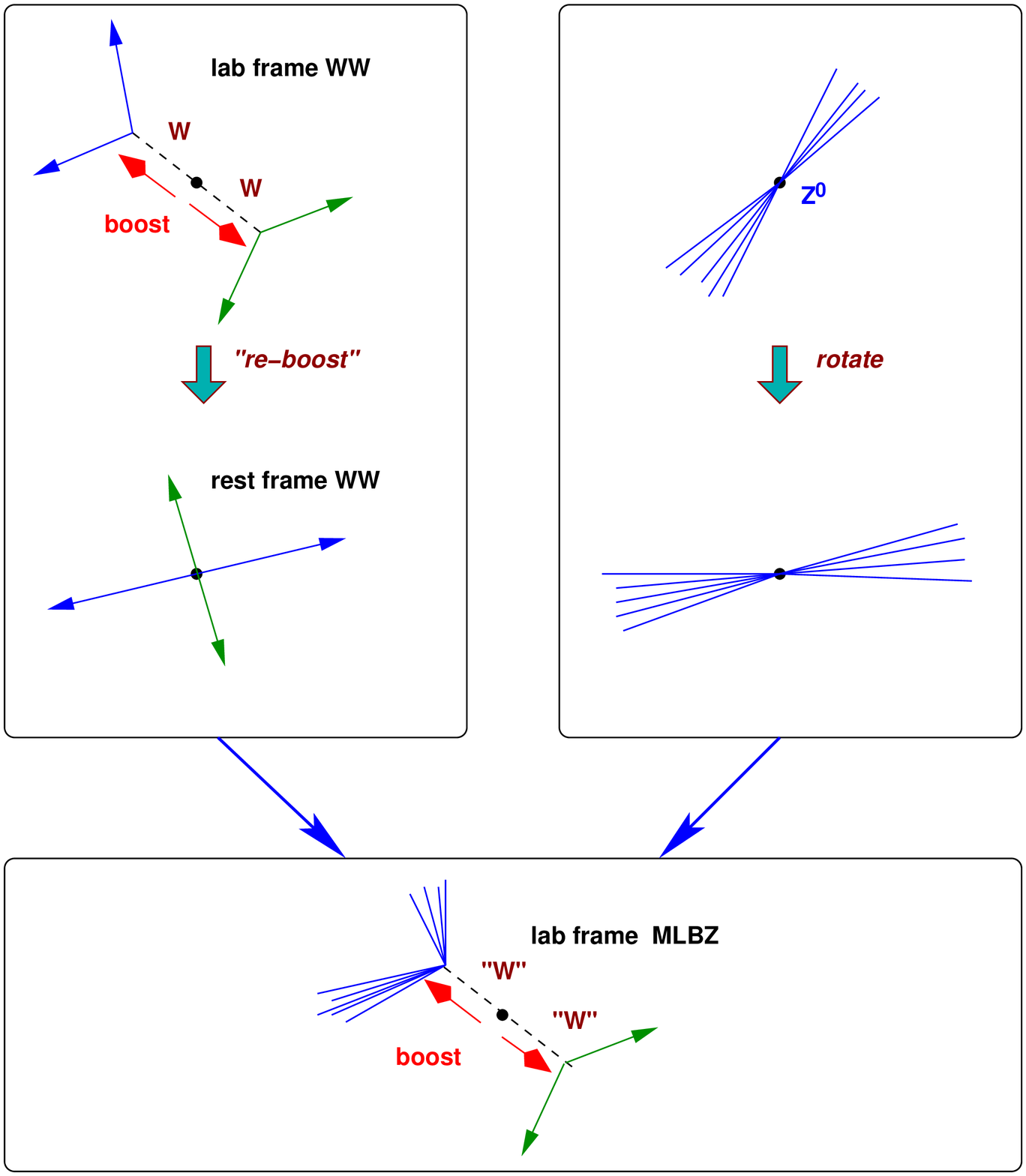}
\vspace{0.5cm}
  \caption{Illustration of the mixing and boosting procedure within the MLBZ
         method (see text for details).}
\efg{mlbzill2}

The main extension of the method beyond that described in~\cite{mlbz} consisted
in an improved mixing and boosting procedure of the $\Z$ events into MLBZs,
demonstrated in Figure~\rf{mlbzill2}.


The 4-momenta of the four primary quarks in \WPHACT\ generated
$\mathrm{W^+W^- \rightarrow \qqqq}$ events were used as event templates.
The $\Z$ events from data or simulation were chosen such that their
thrust axis directions were close
in polar angle to one of the primary quarks of the $\mathrm{W^+W^-}$
event template. Each template W was then boosted to its rest frame.
The particles in the final state of a selected $\Z$ event were
rotated so that the thrust axis matches the rest frame direction of the primary quarks
in the
$\mathrm{W^+W^-}$ template. After rescaling the kinematics of the
$\Z$ events to match the W boson mass in the generated $\mathrm{W^+W^-}$
template, the two $\Z$ events were boosted to the lab frame of the
$\mathrm{W^+W^-}$  template. All particles having
an absolute polar angle with the beam direction smaller than 11$^{\circ}$ were
removed from the event. The same generated \WPHACT\ events were used for the
construction of both the data MLBZs and Monte Carlo MLBZs in order to
increase the correlation between both emulated samples to about 31\%. This
correlation was taken into account when quoting the statistical uncertainty on
the systematic shift on the observables between data and Monte Carlo MLBZs.

It was verified that when introducing a significant mass shift of
300 MeV/c$^2$ on $\mw$ by using the cone rejection algorithm, it was
reproduced
within 15\% by applying the MLBZ technique.
Because the expected systematic uncertainties on the $\Delta \mw(\std,i)$
observables of interest are one order of magnitude smaller than 300 MeV/c$^2$,
this method is clearly justified.

The double difference of Equation~\ref{doublediff} was determined
with the MLBZ method using $\Z$ events selected in the data sets collected
during the 1998 calibration runs and $\Z$ events from the corresponding Monte
Carlo samples. The following results were obtained for the
$\dmwrcn$ observable:

\begin{equation}
\begin{array}{lcrcr}
\Delta_{\mathrm{syst}}({\mathrm{\ARIADNE\ ,DA}})&=& -1.9 &\pm & 3.9\mathrm{(stat)} {\MeVm} \\
\Delta_{\mathrm{syst}}({\mathrm{\PYTHIA\ ,DA}}) &=& -5.7 &\pm & 3.9\mathrm{(stat)} {\MeVm} \\
\Delta_{\mathrm{syst}}({\mathrm{\HERWIG\ ,DA}}) &=&-10.6 &\pm & 3.9\mathrm{(stat)} {\MeVm}
\end{array}
\end{equation}

\noindent
where the statistical uncertainty takes into account the correlation between
the Monte Carlo and the
data MLBZ events, together with the correlation between the two
$\mw$ estimators. This indicates that most of the fragmentation,
detector and Between-W Bose-Einstein Correlation systematics are small.
The study was not performed for the $\dmwp$
observable.

Other systematic sources on the reconstructed jets are not
considered as the $\mw$ estimators used in the difference $\Delta \mw(\std,i)$
have a large correlation.

\subsubsection{Additional Fragmentation systematic study}

The fragmentation of the primary partons is modelled in the Monte Carlo
simulation used for the calibration of the $\mw^i$ observables.

\bfg{htbp}
  \includegraphics[height=16cm]{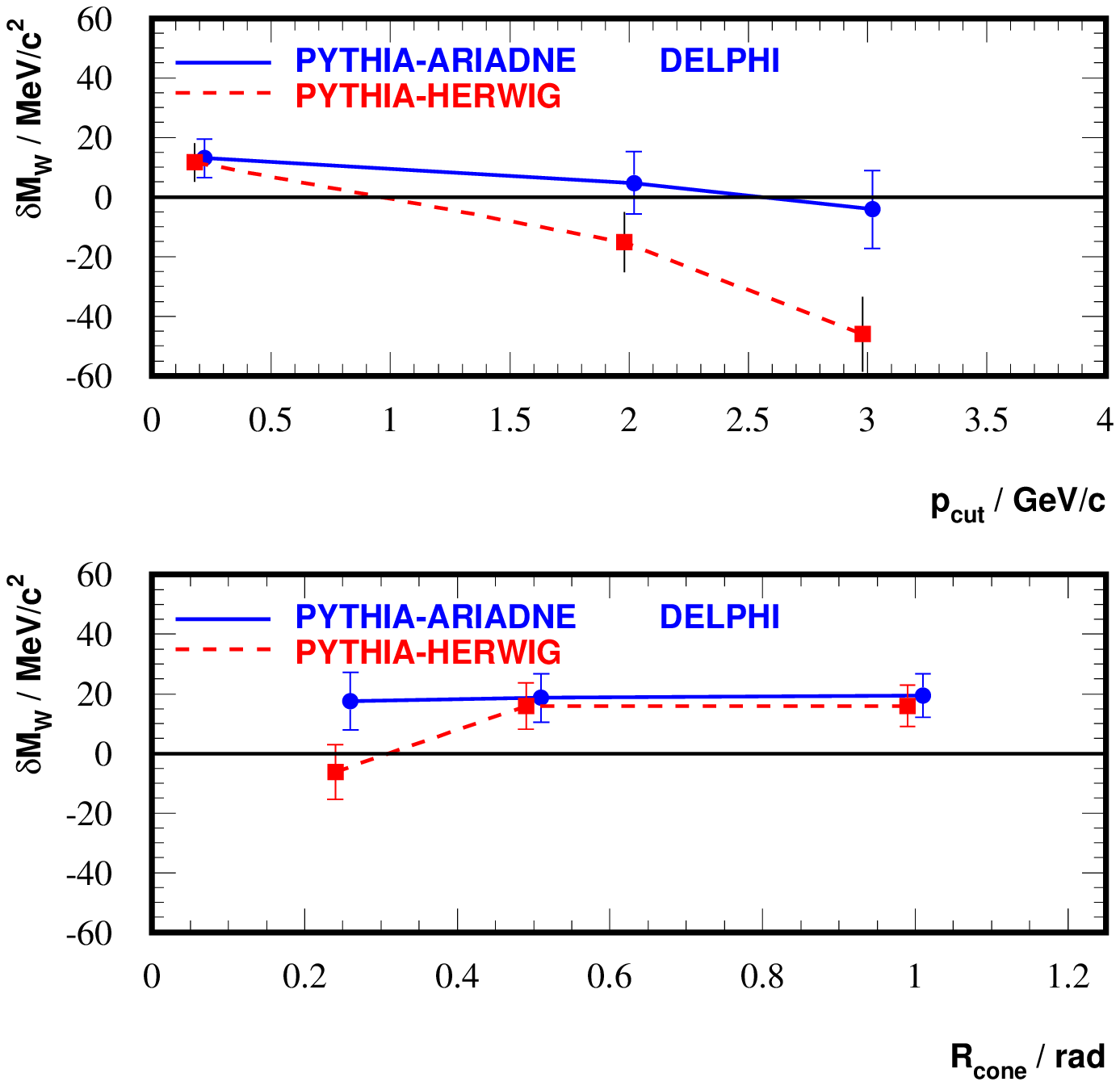}
  \caption{Systematic shifts $\delta \mw$, 
  on $\mw$ observables,
when applying different
fragmentation models as a function of the $\Pcut$ or $\Rcone$
values used in
the construction of the $\mw$ observable.
These Monte Carlo estimates were obtained at a centre-of-mass energy
of 189 GeV. The uncertainties are determined with the Jackknife method.}
\efg{fragsyst}

The expected values on the $\mw$ estimators from simulation
(in the $\mathrm{\kappa=0}$
hypothesis) are changed when using
different fragmentation models~\cite{paper189}, resulting
in systematic uncertainties on the measured $\mw^{i}$ observables
and hence possibly also on our estimated $\mathrm{\kappa}$.
In Figure~\rf{fragsyst} the systematic shift $\delta\mw$ in the different
$\mw^i$ observables is shown when using \HERWIG\ or \ARIADNE\ rather
than \PYTHIA\ as the fragmentation model in the no Colour Reconnection
hypothesis. When inferring $\mathrm{\kappa}$ from the data difference,
$\Delta \mw(\std,i)$, the \PYTHIA\ model is used to
calibrate each $\mw^i$ observable. This data difference for
$\mwpct$,
$\dmwpct$,
changes\footnote{
This change, $\dmwpct^{\PYTHIA} -
\dmwpct^{\HERWIG}$, is
given by\break $\mathrm{\delta M_W(std \equiv p_{cut}=0.2 \, GeV/c)^{{\PYTHIA}-\HERWIG} -
\delta M_W(p_{cut}=2 \, GeV/c)^{{\PYTHIA}-\HERWIG}}$,
and similar expressions for the \ARIADNE\ and $\Rcone$ cases
(for $\Rcone$, $\std\equiv\Rcone=\pi$).}
by (27~$\mathrm{\pm}$~12)~MeV/$\mathrm{c^2}$ or
(8~$\mathrm{\pm}$~12)~MeV/$\mathrm{c^2}$ when
replacing \PYTHIA\ by respectively \HERWIG\ or {\ARIADNE}.
Similarly, the observable $\dmwrcn$ changes by
(-4 $\mathrm{\pm}$ 10) MeV/$\mathrm{c^2}$
or (-6 $\mathrm{\pm}$ 10) MeV/$\mathrm{c^2}$ when replacing
\PYTHIA\ by respectively \HERWIG\ or {\ARIADNE}.
The largest shift of the observable when 
changing 
fragmentation models
(or the uncertainty on this shift if larger) 
is taken as systematic uncertainty on 
the 
value of the
observable. 
Hence, systematic errors of 27 MeV/$\mathrm{c^2}$
for the $\dmwpct$ observable
and 10 MeV/$\mathrm{c^2}$ for the
$\dmwrcn$ observable were assumed
as the contribution from fragmentation uncertainties
.
The MLBZ studies (see above) are compatible with these results,
hence no additional systematic due to fragmentation
was quoted for the $\dmwrcn$ observable.

\subsubsection{Energy Dependence}

The biases of the different $\mw$ estimators have a different
dependence on the centre-of-mass energy,
hence the calibration of $\Delta \mw(i,j)$
will be energy dependent.
The energy dependence of each individual $\mw$ estimator was
parameterised with a second order polynomial.
Since WPHACT event samples were used at a range of centre-of-mass
energies the uncertainty on the parameters describing these
curves are small.
Therefore a small systematic uncertainty of 3 MeV/c$^2$ was quoted on the
$\Delta \mw(i,j)$ observables due to the calibration.

\subsubsection{Background}

The same event selection criteria were applied for all the $\mw$ estimators, hence the same
background contamination is present in all analyses.
The influence of the $\mathrm{q{\bar q}(\gamma)}$ background events on the
individual
$\mw$ estimators is small~\cite{paper189} and was taken into
account
when constructing the centre-of-mass energy dependent calibration curves of
the individual
$\mw$ estimators. The residual systematic uncertainty on both
$\Delta \mw(i,j)$ observables is 3 MeV/c$^2$.

\subsubsection{Bose-Einstein Correlations}


As for the particle flow method, the systematic uncertainties due to possible
Bose-Einstein Correlations are estimated via Monte Carlo simulations.
The relevant values for the systematic uncertainties on the observables are the
differences between the observables obtained from the Monte Carlo events with
Bose-Einstein Correlations inside individual W's (BEI) and those
with, in addition, Bose-Einstein Correlations between identical particles from
different W's (BEA). The values were
estimated to be (6.4 $\mathrm{\pm}$ 9.3) MeV/$\mathrm{c^2}$ for
the $\dmwpct$ observable, and
(7.2 $\mathrm{\pm}$ 8.2) MeV/$\mathrm{c^2}$ for
the $\dmwrcn$ observable.
As the uncertainties in the estimated contributions were larger than the contributions
themselves, these uncertainties were added in quadrature to the systematic errors on
the relevant observables.

\subsubsection{Cross-check in the Semi-leptonic Channel}

Colour Reconnection between the decay products
originating from different W boson decays can only occur in
the $\mathrm{W^+W^- \rightarrow \qqqq}$ channel. The semi-leptonic
$\mathrm{W^+W^-}$ decay channel (i.e, $\qqb\len$) is by definition free of those effects.
Therefore the determination of Colour Reconnection sensitive observables,
like $\dmwrcn$, in this decay channel
could indicate the possible presence of residual systematic effects.
A study of the $\dmwrcn$ observable
was performed in the semi-leptonic decay channel.
The semi-leptonic $\mw$ analysis in~\cite{paper189} was used and the cone
algorithm was implemented in a similar way as for the fully hadronic decay
channel. The same data sets have been used as presented throughout this paper
and the following result was obtained:

\begin{equation}
\begin{array}{lclclcrcl}
\Delta \mw(\std,\Rcone) &=& \mw^{\std}     &-& \mw^{\Rcone}    &=&
  (8  &\pm& 56 \mathrm{(stat)})\MeVm \\
\end{array}
\end{equation}

\noindent
where the statistical uncertainty was 
computed taking into account the
correlation between both measurements. Although the statistical significance
of this cross-check is small, a good agreement was found for
both $\mw$ estimators.

\subsection{Results from the \boldmath{$\mw$} Estimators Analyses}
\label{section:results}

The observable $\dmwr$
with $\Rcone$ equal to 0.5 rad (defined above),
was found to be the most sensitive to 
the \SKI\ Colour Reconnection model, and
the $\dmwpct$ observable was
measured as a cross-check.
The analyses were calibrated with
\PYTHIA\ $\mathrm{\kappa=0}$ \WPHACT\ generated simulation events.
The values measured from the combined DELPHI data at centre-of-mass
energies ranging between
183
and 208 GeV are:

\begin{equation}
\begin{array}{lclclcrcrcl}
\dmwr &=& \mw^{\std}     &-& \mwr &=&  (59
&\pm& 35\mathrm{(stat)}  &\pm& 14\mathrm{(syst)})\MeVm \\
\dmwp &=& \mw^{\std}     &-& \mwp &=& (143
&\pm& 61\mathrm{(stat)}  &\pm& 29\mathrm{(syst)})\MeVm \\
\end{array}
\end{equation}

\noindent
where the first uncertainty numbers represent the statistical components and the second the
combined systematic ones.
The full breakdown of the uncertainties
on both observables can be found in Table~\ref{table:syst2}.

\begin{table}[h]
 \begin{center}
  \begin{tabular}{|l|c|c|}
\hline
                    & \multicolumn{2}{|c|}{Uncertainty contribution (MeV/c$^2$)} \\
Source              & $\dmwrcn$
                    & $\dmwpct$             \\
\hline
\hline
Fragmentation                                         &  11   &  27      \\
Calibration                                           &   3   &   3      \\
Background                                            &   3   &   3      \\
BEI-BEA                                               &   8   &   9      \\
\hline
Total systematic                                      &  14   &  29      \\
\hline
Statistical Error                                     &  35   &  61      \\
\hline
\hline
Total                                                 &  38   &  67      \\
\hline
  \end{tabular}
\caption{Breakdown of the total uncertainty on both relevant observables. }
  \label{table:syst2}
 \end{center}
\end{table}



From these values estimates were made for the
$\mathrm{\kappa}$ parameter by comparing them with the
Monte Carlo expected values in different hypothesis of
$\mathrm{\kappa}$, shown in Figure~\rf{observablekappa} for
the observable $\dmwrcn$.

\bfg{htbp}
  \includegraphics[width=\linewidth]{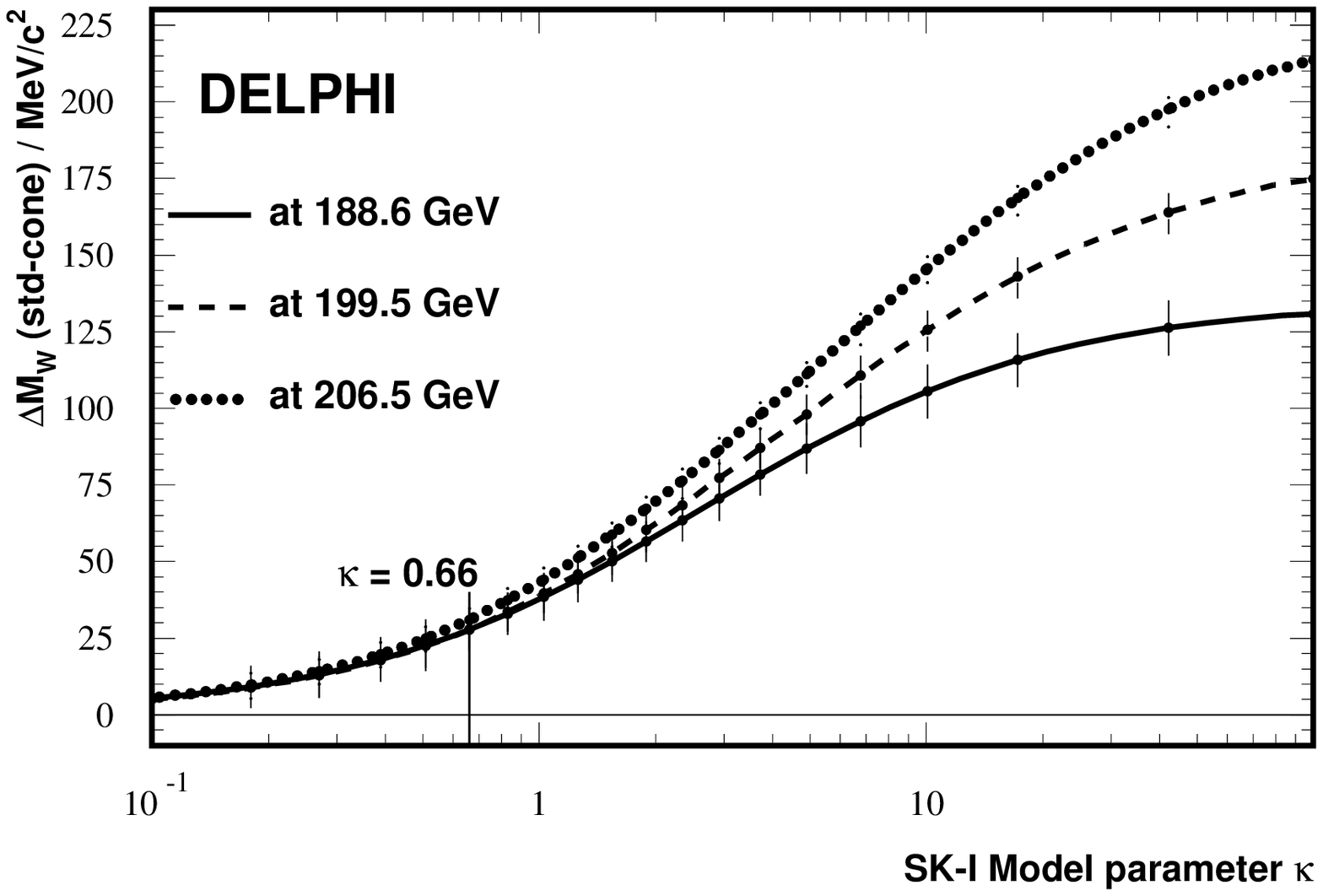}
  \caption{The dependence of the observable $\dmwrcn$ from simulation
  events on the
  value of the \SKI\ model parameter $\kappa$. The dependence is given
  at three centre-of-mass energies.}
\efg{observablekappa}

The Gaussian uncertainty on the measured observables was used to construct
a log-likelihood function
$\mathrm{{\cal L}(\kappa)} = \mathrm{-2 \log L(\kappa)}$ for
$\mathrm{\kappa}$.
The log-likelihood
function obtained is shown in Figure~\rf{likkappa} for the first
and in Figure~\rf{likkappapcut} for the second observable.

\bfg{htbp}
  \includegraphics[height=16cm]{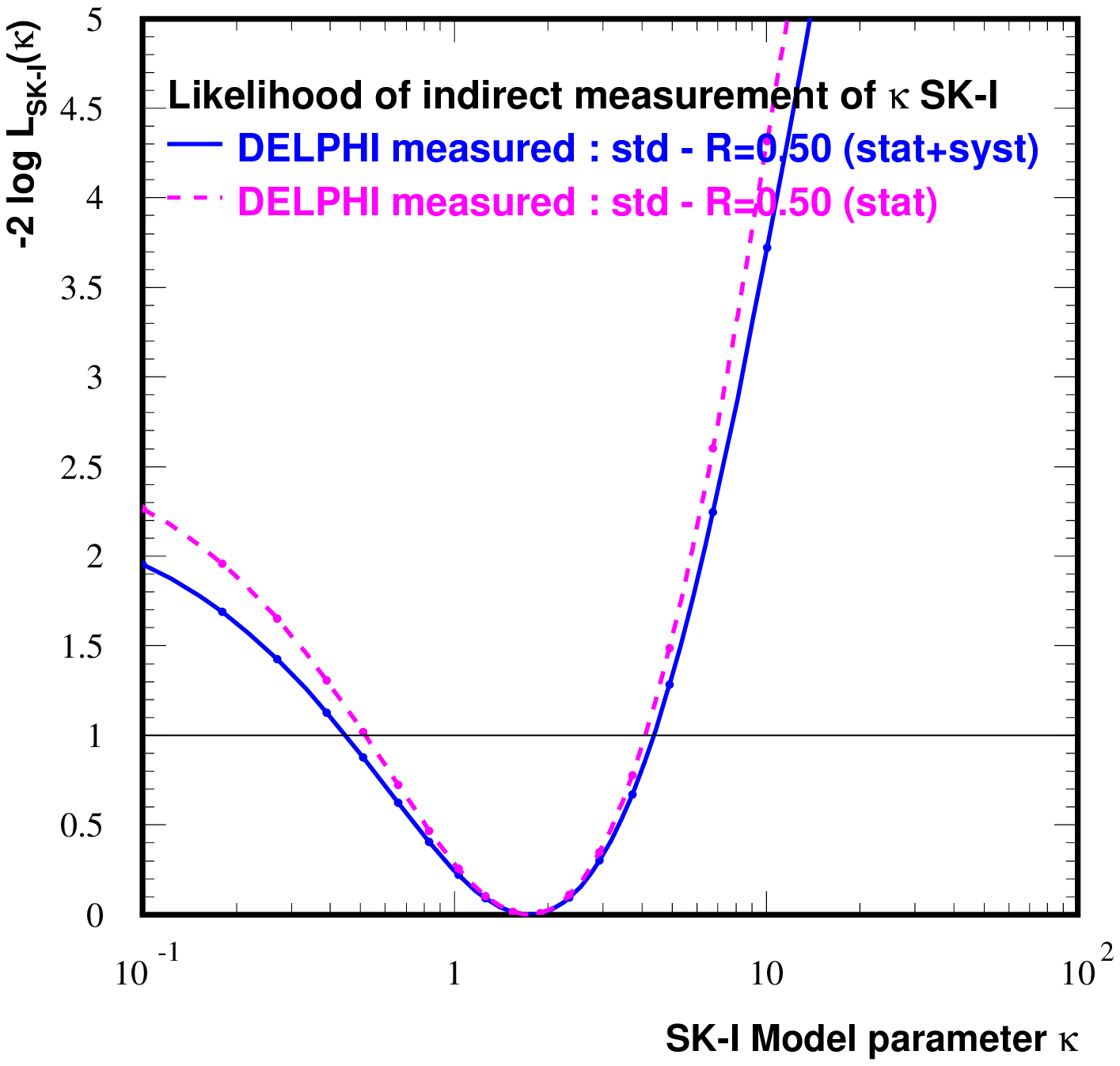}
  \caption{The log-likelihood function $-2\log{\mathrm L}(\kappa)$ obtained from the
DELPHI data measurement of $\dmwrcn$. The 
bottom curve (full line) gives the final result including the statistical
uncertainty on $\dmwrcn$ and the investigated
systematic uncertainty contributions. The 
top curve (dashed) 
is centred on the same minimum and reflects the log-likelihood function
obtained when only statistical uncertainties are taken into account.
}
\efg{likkappa}

\bfg{htbp}
  \includegraphics[height=16cm]{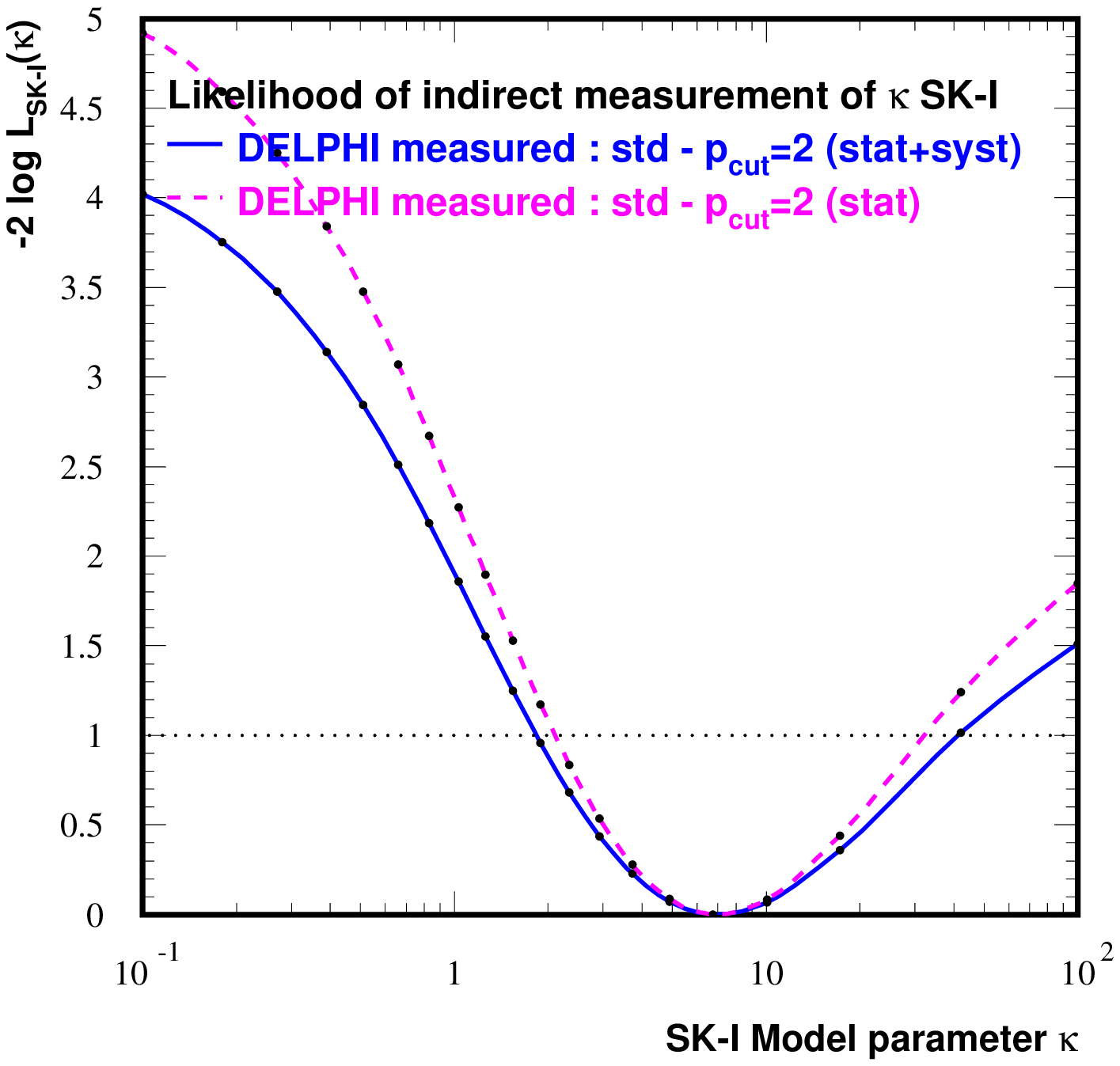}
  \caption{The log-likelihood function $-2\log{\mathrm L}(\kappa)$ obtained
from the 
DELPHI data measurement of $\dmwpct$. The 
bottom curve (full line) gives the final result including the statistical
uncertainty on $\dmwpct$ and the investigated
systematic uncertainty contributions. The 
top curve (dashed) 
is centred on the same minimum and reflects the log-likelihood function
obtained when only statistical uncertainties are taken into account.
}
\efg{likkappapcut}

The result shown in Figure~\rf{likkappa}
is the primary result of this analysis,
because of the larger
sensitivity of the $\dmwrcn$ observable
to the value of
$\mathrm{\kappa}$ (see section~\ref{section:meask}).
The value of $\kappa$ most compatible with the data within one standard
deviation of the measurement is
\begin{equation}
\kappa_{\mathrm{\mbox{\SKI}}} = 1.75^{+2.60}_{-1.30}\,\, .
\end{equation}

The result on $\mathrm{\kappa}$ extracted from the
cross-check $\dmwpct$ observable is found
not to differ significantly from the quoted result obtained with the more
optimal $\dmwrcn$ observable.
The significance can be
determined by the difference between both $\mw$ estimators :

\begin{equation}
\begin{array}{lclcrcrl}
\mwp &-& \mwr    &=& (-84 &\pm& 59\mathrm{(stat)})\MeVm & .
\end{array}
\end{equation}

\noindent
Taking into account that the expectation of this difference depends on
$\kappa$, we find a statistical deviation of about 1 to 1.5$\sigma$ between the
measurements. No improved sensitivity is obtained by combining the
information of both observables.

In this paper the \SKI\ model for Colour Reconnection implemented in
\PYTHIA\ was studied because it parameterizes the effect as function of the
model parameter $\mathrm{\kappa}$. Other phenomenological models implemented in
the {\ARIADNE}~\cite{ariadne,CR-GH} and {\HERWIG}~\cite{herwig} Monte Carlo
fragmentation schemes exist and are equally plausible. Unfortunately their
effect in $\mathrm{\WW \rightarrow \qqqq}$ events cannot be scaled with a
model parameter, analogous to $\mathrm{\kappa}$ in {\SKI}, without affecting the
fragmentation model parameters. Despite this non-factorization property,
the consistency of these models with the data can still be examined.
The Monte Carlo predictions of the observables in the hypothesis with
Colour Reconnection (calibrated in the hypothesis of no Colour Reconnection)
give the following values:
\begin{center}
\begin{equation}
\begin{array}{lclclcrcrlc}
{\mathrm \ARIADNE\ } & \rightarrow & \mw^{\std} &-& \mwr &=&  (7.2  &\pm& 4.1) & \MeVm & \\
{\mathrm \ARIADNE\ } & \rightarrow & \mw^{\std} &-& \mwp &=&  (9.4  &\pm& 7.0) & \MeVm & \\
{\mathrm \HERWIG\ }  & \rightarrow & \mw^{\std} &-& \mwr &=& (19.7  &\pm& 4.0) & \MeVm & \\
{\mathrm \HERWIG\ }  & \rightarrow & \mw^{\std} &-& \mwp &=& (22.8  &\pm& 6.9) & \MeVm & . \\
\end{array}
\end{equation}
\end{center}
The small effects on the observables with the \HERWIG\ implementation of
Colour Reconnection compared to those predicted by \SKI\ are due to the fact
that the fraction of events that reconnect is smaller in \HERWIG\
(1/9) compared to \SKI\ ($\gtrsim$ 25\% at $\sqrt{s}=200$~GeV). 
After
applying this scale factor between both models, their predicted
effect on the W mass and on the $\Delta \mw(i,j)$ observables becomes compatible.
The \ARIADNE\ implementation of Colour Reconnection has a much smaller
influence on the observables compared to those predicted with
the \SKI\ and \HERWIG\ Monte Carlo.

\subsection{Correlation with Direct \boldmath{$\mw$} Measurement}
\label{section:correlation}

When using a data observable to estimate systematic uncertainties on some
measurand inferred from the same data sample, the
correlation between the estimator used to measure the systematic bias and the
estimator of the absolute value of the measurand
should be taken into account. Therefore the
correlation between the Colour Reconnection sensitive observables
$\dmwrcn$ and
$\dmwpct$ and the absolute
$\mw(\std)$ estimator was calculated. The correlation was
determined from the Monte Carlo events and with 
$\mathrm{\kappa = 0}$ or no Colour Reconnection.
The values obtained were found to be stable as a function of $\mathrm{\kappa}$
within the statistical precision. The correlation between
$\dmwrcn$ and
$\mw(\std)$ was found to be 11\%, while for the one between
$\dmwpct$ and
$\mw(\std)$ a value of 8\% was obtained.
Also the correlation between the different
$\mw$ estimators was estimated and found to be stable
with the value of $\mathrm{\kappa}$.
A value of 83\% was obtained for the correlation between $\mw(\std)$
and $\mwrcn$, while 66\% was obtained between
$\mw(\std)$ and $\mwpct$.


\section{Combination of the Results in the Scope of the \SKI\ Model}

%

The log-likelihood curve from the particle flow method 
 was
combined with the curve from the $\Delta\mw$ method and the result is
shown in Figure~\rf{likkappacomb}. The correlations between the
analyses were neglected because the overlap between the samples is small
and the nature of the analyses is very different.
The total errors were used (statistical and systematic added in quadrature)
in the combination.

\bfg{htbp}
  \includegraphics[height=16cm]{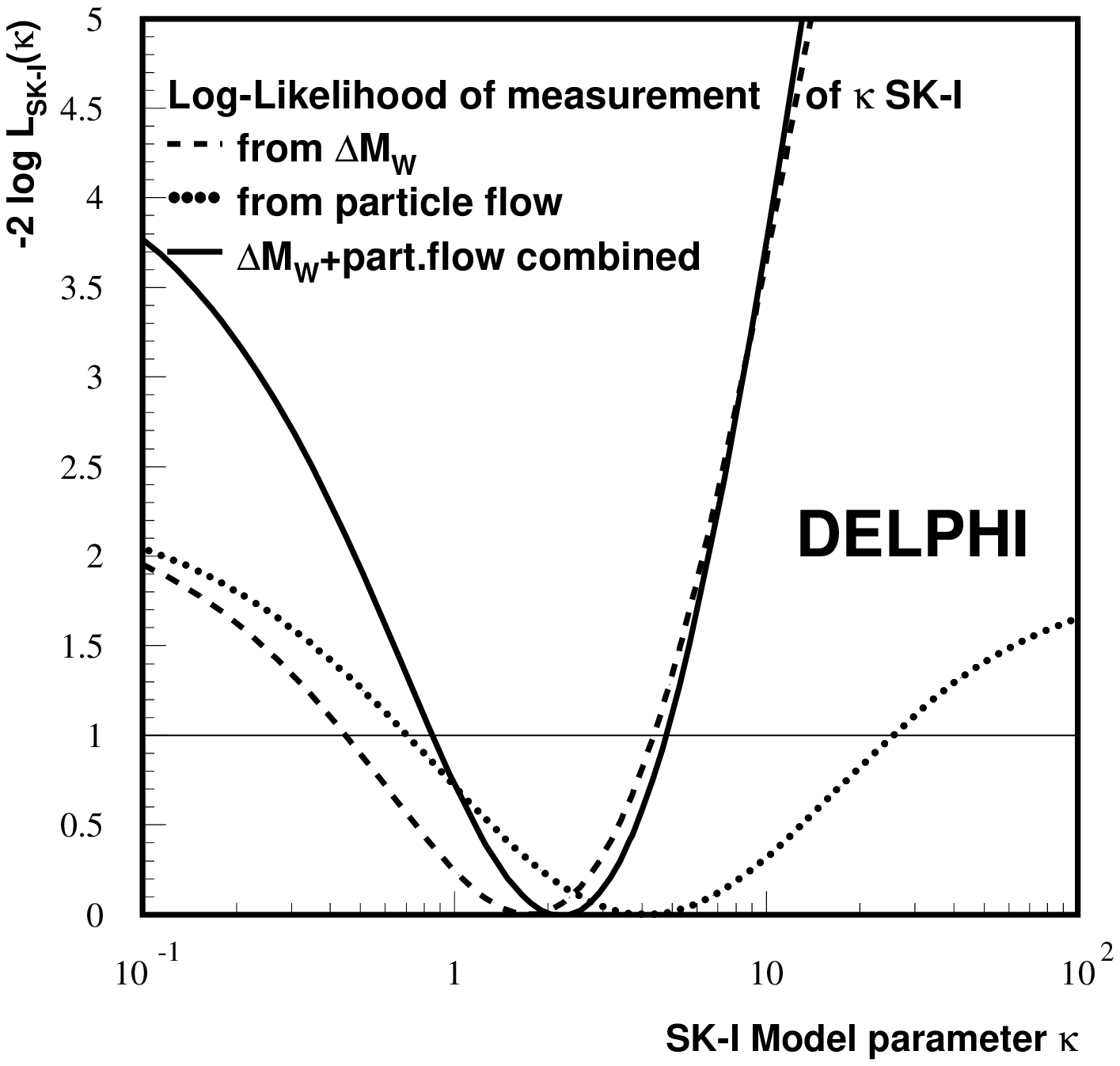}
  \caption{The log-likelihood
function $-2\log{\mathrm L}(\kappa)$ obtained from the combined
DELPHI measurement via $\dmwrcn$ and the
particle flow. The full line 
gives the final result including
the statistical and systematic uncertainties. The log-likelihood
functions are combined in the hypothesis of no correlation between the
statistical and systematic uncertainties of both measurements.}
\efg{likkappacomb}


The best value for $\kappa$ from the minimum of the curve,
with its error given by the width of the curve at the value
$-2\log {\mathrm L}=(-2\log {\mathrm L})_{\mathrm{min}}+1$, is:
\begin{equation}
\kappa_{\mathrm{\mbox{\SKI}}} = 2.2^{+2.5}_{-1.3}\, .
\end{equation}

\section{Conclusions}

Colour Reconnection (CR) effects in the fully hadronic decays of W pairs,
produced in the DELPHI experiment at LEP, were investigated using the
methods of the particle flow and the $\mw$ estimators, notably the
$\dmwrcn$ observable.


The average of the ratios $R$ of the integrals between 0.2 and 0.8 of the
particle distribution in Inside-W regions to the Between-W regions was found to be

\begin{equation}
\langle R\rangle = 0.979 \pm 0.032 \mathrm{(stat)} \pm 0.035
\mathrm{(syst)} \, .
\end{equation}

\noindent The values used in this average were obtained
after rescaling the value at each energy to the value at
a centre-of-mass energy of 189 GeV
using a fit to the MC without CR.

The effects of CR on the values of the reconstructed mass of the W boson,
as implemented in different Monte Carlo models, were studied with different
estimators. From the estimator of the W mass with the strongest sensitivity
to the \SKI\ model of CR, the
$\dmwrcn$ method,
the difference in data was found to be
\begin{equation}
\dmwr=
\mathrm{\mw^{\std}}-\mwrcn = (\, 59 \pm 35\mathrm{(stat)} \pm 14\mathrm{(syst)}\, )\MeVm \, .
\end{equation}

From the combination of the results from particle flow and $\mw$ estimators,
corresponding to the curve in full line
shown in Figure~\rf{likkappacomb}, the best value and total error
for the $\kappa$ parameter in the \SKI\ model was extracted to be:
\begin{equation}
\kappa_{\mathrm{\mbox{\SKI}}}= 2.2^{+2.5}_{-1.3}\label{resultkappa}
\end{equation} which
corresponds to a probability of reconnection of
${\cal P}_{\mathrm{reco}}=52\%$ and lies in the range
$\mathrm{31\% < {\cal P}_{reco} < 68\%}$  at 68\% confidence level.

The two analysis methods used in this paper are complementary: the method of
particle flow provides a model-independent measurement
but has significantly less sensitivity towards the \SKI\ model of CR than
the method of $\Delta\mw$ estimators.
%

The obtained value of $\kappa$ in equation~(\ref{resultkappa}) can be compared with similar
values obtained by other LEP experiments, and it was found to be
compatible with, but higher than,
the values obtained with the particle flow by L3~\cite{l3cr} and
OPAL~\cite{opalcr}. It is also compatible with, but higher than, the values
obtained with the method of different $\mw$ estimators by OPAL~\cite{opalmw}
and ALEPH~\cite{alephmw}.

\subsection*{Acknowledgements}
\vskip 3 mm
We thank the
\ALEPH\ Collaboration for the production of the
simulated ``Cetraro Samples''.

We are greatly indebted to our technical 
collaborators, to the members of the CERN-SL Division for the excellent 
performance of the LEP collider, and to the funding agencies for their
support in building and operating the DELPHI detector.\\

We acknowledge in particular the support of \\
Austrian Federal Ministry of Education, Science and Culture,
GZ 616.364/2-III/2a/98, \\
FNRS--FWO, Flanders Institute to encourage scientific and technological 
research in the industry (IWT) and Belgian Federal Office for Scientific, 
Technical and Cultural affairs (OSTC), Belgium, \\
FINEP, CNPq, CAPES, FUJB and FAPERJ, Brazil, \\
Czech Ministry of Industry and Trade, GA CR 202/99/1362,\\
Commission of the European Communities (DG XII), \\
Direction des Sciences de la Mati$\grave{\mbox{\rm e}}$re, CEA, France, \\
Bundesministerium f$\ddot{\mbox{\rm u}}$r Bildung, Wissenschaft, Forschung 
und Technologie, Germany,\\
General Secretariat for Research and Technology, Greece, \\
National Science Foundation (NWO) and Foundation for Research on Matter (FOM),
The Netherlands, \\
Norwegian Research Council,  \\
State Committee for Scientific Research, Poland, SPUB-M/CERN/PO3/DZ296/2000,
SPUB-M/CERN/PO3/DZ297/2000, 2P03B 104 19 and 2P03B 69 23(2002-2004)\\
FCT - Funda\c{c}\~ao para a Ci\^encia e Tecnologia, Portugal, \\
Vedecka grantova agentura MS SR, Slovakia, Nr. 95/5195/134, \\
Ministry of Science and Technology of the Republic of Slovenia, \\
CICYT, Spain, AEN99-0950 and AEN99-0761,  \\
The Swedish Research Council,      \\
Particle Physics and Astronomy Research Council, UK, \\
Department of Energy, USA, DE-FG02-01ER41155, \\
EEC RTN contract HPRN-CT-00292-2002. \\


\pagebreak

\end{document}